\pdfoutput=1
\documentclass{article}
\usepackage[a4paper, total={6in, 8in}]{geometry}
\usepackage{xcolor}
\usepackage[title,titletoc]{appendix}
\usepackage[colorlinks,
            citecolor={blue!55!black},
            linkcolor={blue!55!black},
            urlcolor={blue!55!black},
            pdfborder={0 0 0},
            bookmarks=true,
            bookmarksnumbered=true,
            breaklinks=true]{hyperref}
\usepackage{graphicx} 
\usepackage{authblk}  
\usepackage{amsmath} 
\usepackage[capitalize, noabbrev]{cleveref}
\crefname{section}{Section}{Sections}
\crefname{subsection}{Section}{Sections}
\crefname{figure}{Figure}{Figures}
\crefname{table}{Table}{Tables}
\crefname{equation}{Eq.}{Eqs.}
\usepackage{booktabs}
\usepackage[labelformat=simple,font=small,labelfont=bf]{subcaption}

\usepackage{amssymb}
\usepackage{pifont}
\usepackage{makecell}
\usepackage{multicol}
\usepackage[most]{tcolorbox}
\usepackage[numbers,sort&compress]{natbib}
\usepackage{listings}
\usepackage{microtype}          
\newcommand{\instrMCP}{\textsc{instrMCP}}
\newcommand{\qdevBench}{\textsc{q-devBench}}
\newcommand{\QCoDeS}{\textsc{QCoDeS}}
\newcommand{\QMeasure}{\textsc{QMeasure}}

\newcommand{\SQLite}{\textsc{SQLite}}
\newcommand{\bdash}{\hskip0pt\textendash\hskip0pt}

\title{Evaluating Instrument-Centered AI Agents for Table-Top Science Experiments}

\author[1,*,$\dagger$]{Jiaqi Cai}
\author[3,*]{Siyu Cheng}
\author[1]{William S. Huxter}
\author[1]{Li-Qiao Xia}
\author[1]{Shuwen Sun}
\author[2]{Shuai Yuan}
\author[2]{Yifan Zhao}
\author[2]{Naiyuan Zhang}
\author[2]{Weijie Li}
\author[3]{Muxian Xu}
\author[3]{Ilija Zeljkovic}
\author[2,4]{Xiaodong Xu}
\author[1,$\dagger$]{Pablo Jarillo-Herrero}
\affil[1]{Department of Physics, Massachusetts Institute of Technology, Cambridge, MA 02139, USA}
\affil[2]{Department of Physics, University of Washington, Seattle, WA 98195, USA}
\affil[3]{Department of Physics, Boston College, Chestnut Hill, MA 02467, USA}
\affil[4]{Department of Materials Science and Engineering, University of Washington, Seattle, WA 98195, USA}
\affil[*]{These authors contributed equally to this work.}
\affil[$\dagger$]{Corresponding authors: jiaqic@mit.edu; pjarillo@mit.edu}

\date{\today}

\begin{document}

\maketitle

\begin{abstract}
Reliable evaluation of AI agents for table-top science experiments has
remained elusive because live instruments are hard to reset and
share at scale, while instrument statefulness and safety constraints complicate evaluation.
We introduce \instrMCP\ and \qdevBench, an evaluation stack for
instrument-centered AI agents in table-top experiments.
\instrMCP\ connects coding agents to scientific instruments through
the Model Context Protocol, serving as the instrument-access harness.
\qdevBench\ reconstructs 37 real experimental tasks from
condensed-matter physics as virtual
\QCoDeS\ instruments covering quantum electronic transport,
scanning probe imaging, and calibration workflows.  \qdevBench\
exposes reproducible failure modes in laboratory-agent systems while
preserving realistic instrument-control workflows.
Across a range of commercial coding-agent configurations,
current systems already show substantial operational competence,
satisfying 70--83\% of measurement-related rubric criteria.  The main bottleneck is scientific analysis, where performance falls
to 28--46\%.  In an analysis-only ablation in which agents receive clean reference datasets without running the virtual measurement session, analysis performance improves substantially, closing roughly two thirds of the difference between the measurement and analysis scores. This suggests that much of the deficit arises from the measurement-session context, including tool-use overhead, longer context, measurement-planning decisions, and possible acquisition errors, while a residual scientific-reasoning gap remains.
Together, \instrMCP\ and \qdevBench\ provide a practical sandbox
for benchmarking, diagnosing, and improving laboratory AI harnesses,
and establish a route from controlled virtual evaluation to future
live-hardware validation.
\end{abstract}

\section{Introduction}

Table-top experiments have produced many defining
discoveries of modern physics.  However, their efficiency is now limited by the large size of the
experimental parameter space.  In condensed-matter physics, modern devices combine many tunable
parameters with several independent measurement channels, producing
high-dimensional parameter spaces in which signals of interest are
sparse.  Recent examples include
gate-tunable phase diagrams in
twistronics~\cite{cao2018correlated,cao2018unconventional,yankowitz2019tuning,sharpe2019emergent,serlin2020intrinsic,mak2022semiconductor,cai2023signatures,zeng2023thermodynamic,park2023observation,xu2023observation,lu2024fractional},
non-equilibrium optical control of quantum
phases~\cite{basov2017towards,fausti2011light,mitrano2016possible,holtzmann2026optical,cai2026optical,huber2026optical},
and quantum simulation and computation
experiments~\cite{georgescu2014quantum,bloch2008many,chiu2025continuous,manetsch2025tweezer,iqbal2024non,mazurenko2017cold,zwolak2023colloquium}.
Sampling these spaces requires sophisticated automation, but standard
industrial automation is ill-suited to the task~\cite{huang2013software}, as
scientific instrumentation is a long-tail application space of
rapidly evolving, custom setups.  The community has responded with
code-first instrument-control frameworks such as
\QCoDeS~\cite{nielsen2025qcodes},
ARTIQ~\cite{Bourdeauducq2021ARTIQ}, and
others~\cite{starkey2013scripted,weber2021pymodaq,koerner2019python},
which seek to standardize drivers and data flow.  Measurements within
such frameworks nevertheless follow scripted sequences: between runs
the human experimentalist stops, inspects the data, and rewrites the control
code by hand.  AI-driven closed-loop control has been demonstrated
across experimental science, from chemistry and materials synthesis to
accelerators, quantum experiments, and plasma control~\cite{king2009automation,coley2019robotic,gao2022autonomous,kusne2020closed,ament2021autonomous,roussel2021turn,melnikov2018active,degrave2022magnetic}.
However, most such systems remain task-specific controllers rather than general experimental agents.

AI agents combine large language models with tool use in a feedback
loop~\cite{wang2024survey,yao2022react,schick2023toolformer}, and, in principle, can replace that manual control flow with multi-step
reasoning over natural-language
goals~\cite{boiko2023autonomous,sim2024chemos,white2023assessment}.
Doing so demands two competences at once: operating the instruments
and reasoning about the data they produce.
Similar systems already execute end-to-end coding tasks in software
engineering~\cite{jimenez2023swe,yang2024swe}. The
laboratory is a harder setting.  Instruments maintain their states across runs, their
responses are subject to drift and noise, and erroneous commands
can damage samples or hardware.  Evaluation is also harder than in
software because live hardware cannot be reset or replicated at scale.  Existing
science-agent benchmarks mainly target computational or
synthesis-planning workflows~\cite{chen2025scienceagentbench,huang2024mlagentbench,boiko2023autonomous,bran2024chemcrow}.
Laboratory deployment also depends strongly on the runtime harness
around the foundation model, which is the component users actually
interact with~\cite{stach2021autonomous,zhu2022all}.

Here we introduce an evaluation stack for instrument-centered AI agents in table-top science experiments. The stack has two components: (1)~\instrMCP, which provides notebook-mediated access to scientific instruments through the Model Context Protocol, with consent-gated tool use for live deployment; and (2)~\qdevBench, which comprises 37 tasks reconstructed from real measurements spanning quantum electronic transport, scanning probe imaging, and calibration workflows in quantum materials including twisted graphene heterostructures, topological magnets, and superconductors. While the stack applies to any experiment in this class, we focus on condensed matter because it is complicated enough to stress both competences. For benchmarking, \qdevBench\ generates sandboxed virtual instruments that replay recorded device responses by mostly interpolating high-dimensional datasets pre-recorded from real experiments. These do not yet reproduce every live-experiment effect, such as drift and hysteresis. We evaluate 10 agent configurations on three commercial platforms (Claude Code, Codex CLI, and Gemini CLI) across 1,850 runs. Rubric decomposition reveals a measurement–analysis gap: agents satisfy 70–83\% of measurement criteria but only 28–46\% of analysis criteria. An analysis-only ablation in which agents receive pre-collected reference data recovers roughly two thirds of this gap, suggesting that much of the deficit arises from the additional harness and contextual complexity of the full measurement session. A residual scientific-reasoning gap remains. Finally, we demonstrate a contribution pipeline which generates tasks from published condensed-matter papers and, eventually, routine experimental logs, supporting an expandable benchmark suite for experimental science.

\section{Benchmark and evaluation setup}\label{sec:setup}

\subsection{The \instrMCP\ framework}\label{sec:framework}

\begin{figure}
    \centering
    \includegraphics[width=\linewidth,height=0.79\textheight,keepaspectratio]{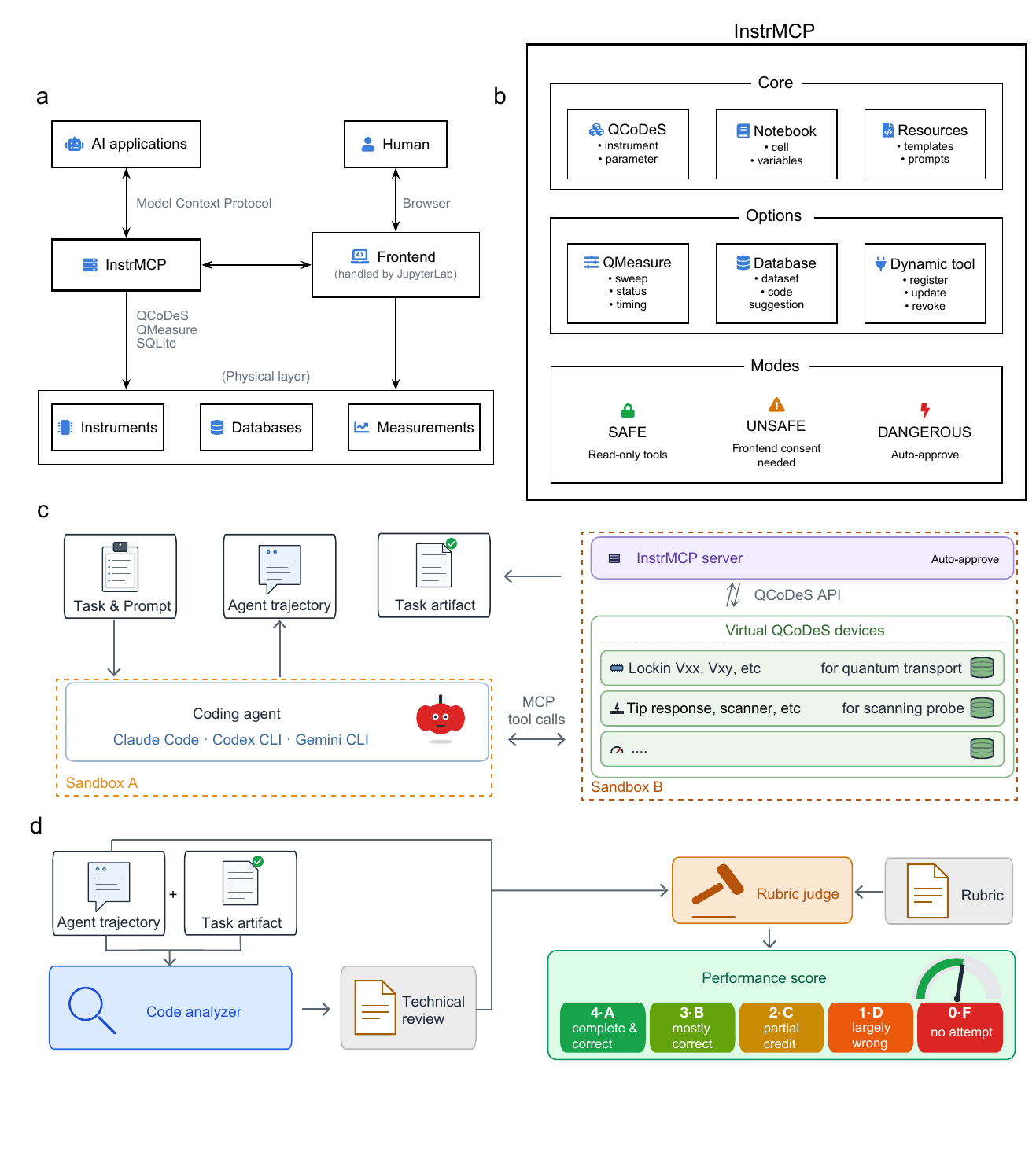}
    \caption{The \instrMCP\ framework and the \qdevBench\ evaluation
    pipeline.
    \textbf{a)}~Module-level view of AI and human connections to a
    shared experiment. \textbf{b)}~Breakdown of the \instrMCP\ server.
    \textbf{c)}~Benchmark runtime. The coding agent (Claude~Code,
    Codex~CLI, or Gemini~CLI) runs in a controlled runtime
    (Sandbox~A) and reaches the virtual \QCoDeS\ devices in a separate sandbox
    (Sandbox~B) only through MCP tool calls. The task prompt enters
    Sandbox~A, and the run produces an agent trajectory and a
    structured task artifact. \textbf{d)}~Two-stage evaluation. A
    code-analyzer LLM reviews the reconstructed code from the
    trajectory and emits a technical review; a rubric-judge LLM then
    scores the task artifact against the rubric on the 0--4 scale
    (see Table~\ref{tab:grading_scale}). Agent-identifying information is
    anonymized before grading.}
    \label{fig:Figure_instrmcp_overview}
\end{figure}

\instrMCP\ connects AI coding agents to scientific instruments through the Model Context Protocol (MCP).  At the system level, Figure~\ref{fig:Figure_instrmcp_overview}a shows the shared experimental workspace in which the AI agent and the human experimentalist connect to the physical layer through \instrMCP\ and frontend (JupyterLab). Figure~\ref{fig:Figure_instrmcp_overview}b displays the major components of a \instrMCP\ server.  The server exposes important components such as \QCoDeS\ instrument control, \QMeasure\ measurement orchestration, and \SQLite\ data persistence through MCP tool calls, while a JupyterLab frontend gives the human experimentalist direct notebook editing, real-time visualization, and manual instrument control.  The shared notebook is also the lab-notebook of record, in which agent and human can append markdown entries to the same file.

\label{sec:safety}%
In laboratory deployment, agent-issued instrument writes are mediated by a consent dialog that the experimentalist must approve before the operation proceeds within the JupyterLab frontend, preserving human authority over irreversible actions.  A three-tier privilege model (safe/unsafe/dangerous) selects which classes of tool are exposed, and an independent static code scanner blocks arbitrary-execution and file-system-escape patterns as an additional defensive layer. Full implementation details of safety control are given in Appendix~\ref{app:instruments}, Section~\ref{sec:safety_impl}.

For benchmark evaluation, the interface is used in a virtual-laboratory mode (Figure~\ref{fig:Figure_instrmcp_overview}c). The coding agent runs in a controlled runtime, while the virtual \QCoDeS\ devices run in a separate sandbox and are reachable only through MCP tool calls.  The task prompt enters the agent runtime, and the completed run produces an agent trajectory and a structured task artifact that includes the modified notebook, the code produced, and plots.

The resulting artifact is evaluated through a two-stage grading pipeline (Figure~\ref{fig:Figure_instrmcp_overview}d).  A code-analyzer large language model (LLM) reviews the reconstructed code and execution timeline, and a rubric-judge LLM scores the task artifact against the task rubric after agent-identifying information has been anonymized.  The task set and virtual instruments used in this benchmark mode are described next.

\subsection{Tasks and virtual instruments}\label{sec:overview}

\begin{figure*}
    \centering
    \includegraphics[width=1\linewidth]{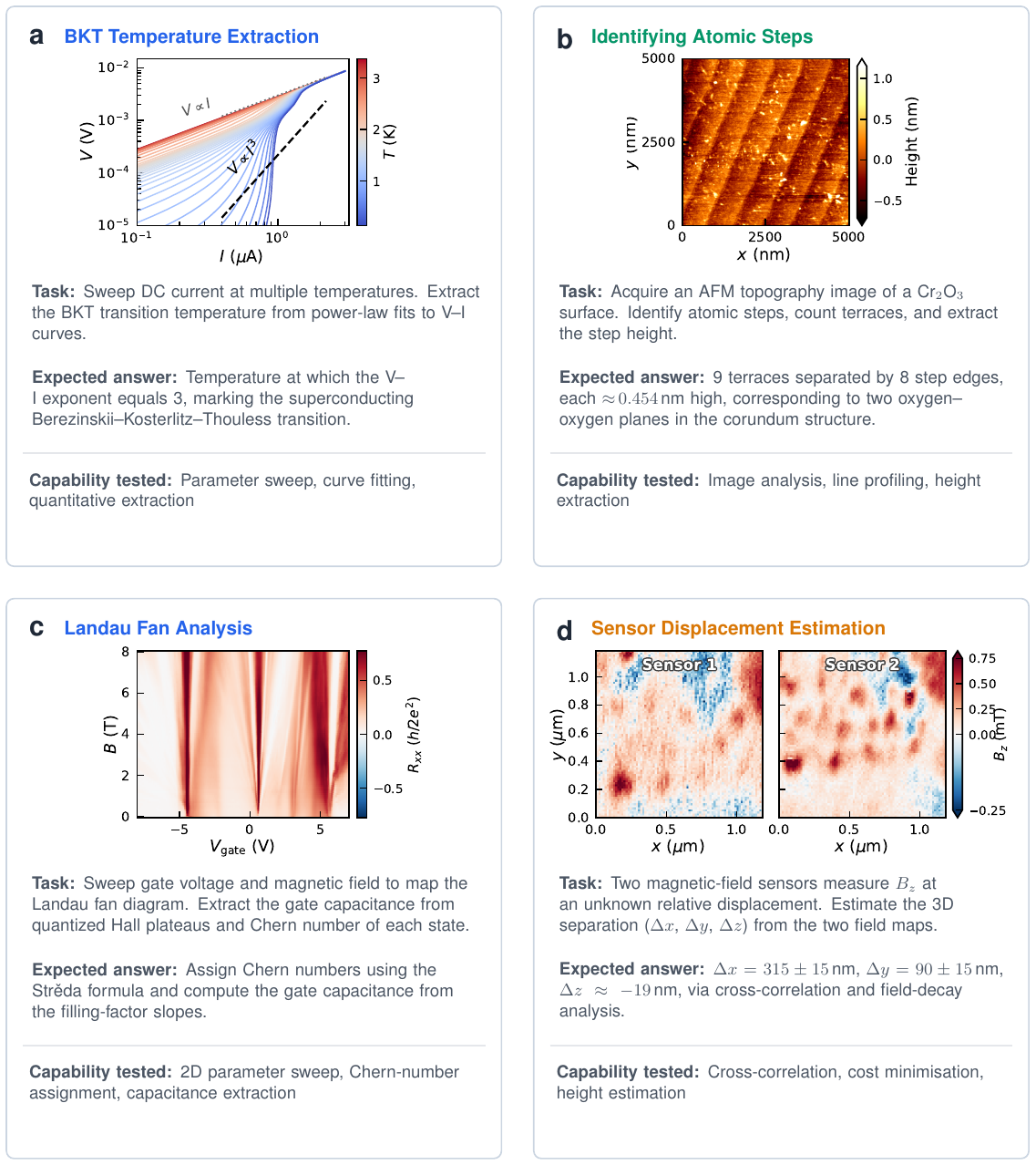}
    \caption{Representative \qdevBench\ tasks.
    \textbf{(a)}~Berezinskii\bdash Kosterlitz\bdash Thouless (BKT) temperature extraction.
    \textbf{(b)}~Identifying atomic steps.
    \textbf{(c)}~Landau fan analysis.
    \textbf{(d)}~Sensor displacement estimation.  Each panel lists the agent capabilities
    tested.}
    \label{fig:task_examples}
\end{figure*}

\qdevBench\ contains 37~tasks drawn from frontier condensed-matter
experiments.  Figure~\ref{fig:task_examples} shows representative tasks from the task set: Berezinskii\bdash Kosterlitz\bdash Thouless (BKT) temperature extraction and Landau-fan analysis illustrate quantum electronic transport tasks (Figure~\ref{fig:task_examples}a,c), while atomic-step identification and sensor displacement illustrate scanning probe imaging tasks (Figure~\ref{fig:task_examples}b,d).  Category counts are summarized in Table~\ref{tab:task_summary}. Tasks are contributed by condensed-matter experimentalists through a dedicated web portal after brief training on the contribution format.  Each contribution includes a natural-language task prompt, a 0--4 grading rubric, annotated raw measurement data, and a source identifier (a citation for published datasets, or the contributor's group and experimental run for private measurements).  Our evaluation prioritizes the correctness and internal consistency of the experimental procedure and reasoning over agreement with a fixed ground-truth outcome.  The full schema is listed in Appendix~\ref{app:benchmark}, Section~\ref{sec:benchmark_design}.

Each task is reconstructed as a virtual \QCoDeS\ instrument.  The generation pipeline infers independent and dependent parameters from the provided data, uses an LLM to produce a \QCoDeS\ driver whose getters interpolate the contributed data, and validates the result by instantiation and parameter-read tests.  At the original sampled points the virtual instruments reproduce the source data exactly for regular-grid tasks. The two exceptions, a large scanning tunneling microscopy (STM) map audited by sub-sampling and the single analytic-model device, are described in Supplemental Section~\ref{supp:instrument_fidelity}.  For selected tasks, task-specific noise is added to mimic experimental imperfections when such noise is important for accomplishing the tasks. 

\begin{table}
\centering
\caption{Distribution of benchmark tasks across experimental categories.
Chern no., Chern number; $T_\mathrm{BKT}$, Berezinskii\bdash Kosterlitz\bdash
Thouless transition temperature; $C_\mathrm{gate}$, gate capacitance;
$\Delta_\mathrm{SC}$, superconducting gap; $H_c$, coercive field;
RSB, rotational symmetry breaking.}
\label{tab:task_summary}
\small
\begin{tabular}{@{} l c l @{}}
\toprule
\textbf{Category} & \textbf{Tasks} & \textbf{Example quantities} \\
\midrule
Quantum electronic transport & 18 & Chern no., $T_\mathrm{BKT}$, $C_\mathrm{gate}$, $\Delta_\mathrm{SC}$, $H_c$ \\
Scanning probe imaging       & 17 & Domain walls, vortices, steps, RSB, displacement \\
Calibration \& model sel.  &  2 & Thermometer calibration, surface charge \\
\midrule
\textbf{Total}             & \textbf{37} & \\
\bottomrule
\end{tabular}
\end{table}

The task set covers 18 quantum electronic transport tasks, 17 scanning probe imaging tasks, and two calibration/model-selection tasks (Table~\ref{tab:task_summary}).  Transport tasks emphasize measurement sweeps and physical-quantity extraction.  Imaging tasks emphasize identifying features in two-dimensional maps, sometimes across different modalities. The calibration/model-selection tasks require more open-ended physical reasoning.  Many tasks require sequential measurements, for which the agent must inspect partial data, decide what is missing, and return to the measurement session before answering.

\subsection{Evaluation protocol}\label{sec:protocol}

Benchmark runs use the sandboxed virtual-laboratory mode introduced above, provided by either the Harbor agent-evaluation framework~\cite{harbor2025} or a custom evaluation stack built on top of \instrMCP.  The \qdevBench-generated virtual \QCoDeS\ drivers replace physical hardware behind the \instrMCP\ interface, and consent is auto-granted via \instrMCP's dangerous mode so that trials run unattended.  During benchmark runs, agents are denied direct access to the raw comma-separated-value (CSV) data files, the hidden task specifications, and the rubric files.  This restriction is enforced in one of two ways: some runs use the Harbor Docker sidecar, which isolates these files at the container boundary, while the remaining runs use a custom-built \instrMCP\ harness that blocks file reads at runtime (Appendix~\ref{app:benchmark}, Section~\ref{sec:agents}).  In both cases, the agent must obtain data through the virtual instruments, so the benchmark measures experimental reasoning rather than file parsing.

We evaluate coding agents as experimental scientists.  Each agent comprises a foundation model, an MCP client that communicates with \instrMCP, and a working memory that maintains context across the session; the typical operational loop and tool-call sequence are described in Appendix~\ref{app:benchmark}, Section~\ref{sec:agents}.  Ten configurations across three commercial agent platforms (Claude~Code, Codex~CLI, and Gemini~CLI) are evaluated with $N=5$ repeated trials per task, totaling 1{,}850 runs. The full configuration list and run settings are given in Appendix~\ref{app:evaluation}, Section~\ref{sec:exp_setup} and Supplemental Table~\ref{tab:model_results}.

Grading follows the two-stage LLM-as-judge pipeline introduced above; implementation details are given in Appendix~\ref{app:evaluation}, Section~\ref{sec:evaluation}.  A stratified second-judge audit of 207 runs found substantial agreement and preserved the ranking extremes (Appendix~\ref{app:evaluation}, Section~\ref{sec:evaluation}; Supplemental Section~\ref{supp:judge_robustness}).

\begin{table}
\centering
\caption{Grading scale used by the LLM judge.}
\label{tab:grading_scale}
\small
\begin{tabular}{@{} c c l @{}}
\toprule
\textbf{Score} & \textbf{Grade} & \textbf{Description} \\
\midrule
4 & A & Complete and correct; all rubric criteria met \\
3 & B & Mostly correct; minor issues or omissions \\
2 & C & Partial credit; significant gaps remain \\
1 & D & Attempted but largely incorrect \\
0 & F & No meaningful attempt or fundamentally wrong \\
\bottomrule
\end{tabular}
\end{table}

\section{Results}\label{sec:results}

\subsection{Overall performance}\label{sec:overall}

\begin{figure}[!htbp]
    \centering
    \includegraphics[width=0.7\columnwidth]{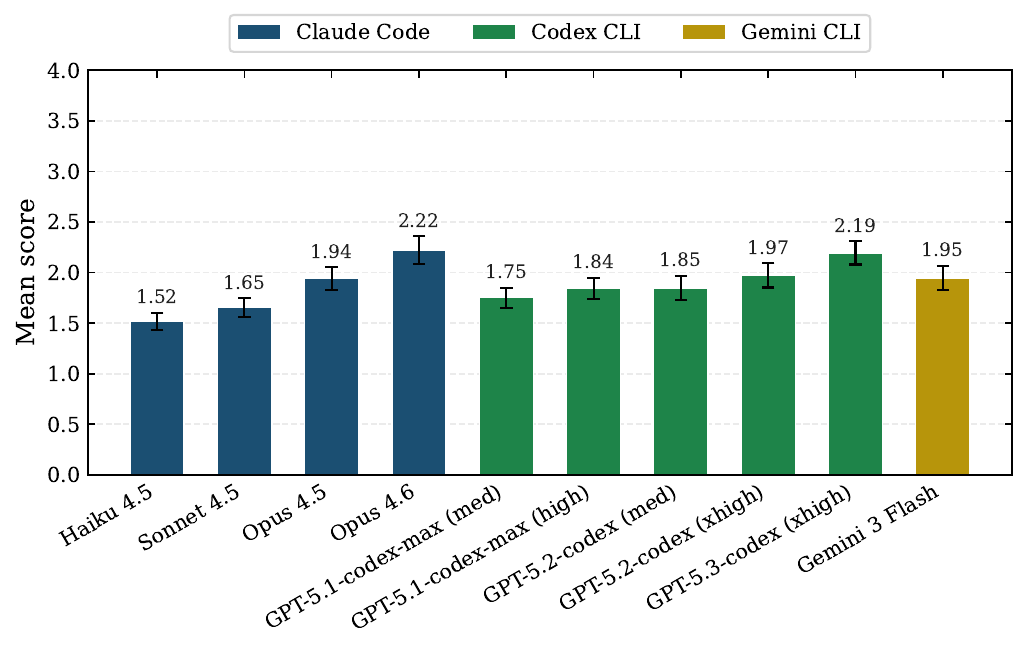}
    \caption{Mean score on the 0--4 rubric scale for each of the ten
    agent configurations.  Error bars are task-clustered
    $\pm 1\sigma_{\text{SEM}}$; 95\% task-clustered bootstrap
    confidence intervals are listed in Supplemental
    Section~\ref{supp:stats}.  Colors indicate agent platform.}
    \label{fig:forest}
\end{figure}

The mean score across all 1{,}850~runs is $1.89$ on the 0--4 rubric. Figure~\ref{fig:forest} shows the per-configuration mean with task-clustered standard-error-of-the-mean (SEM) bars of width $\pm 1\sigma_{\text{SEM}}$.  Claude Code Opus~4.6 ($2.22 \pm 0.14$) and Codex GPT-5.3-codex at extra-high (xhigh) effort ($2.19 \pm 0.11$) achieve the highest mean scores. A middle band around $1.84$--$1.97$ (SEMs $\approx 0.10$--$0.12$) follows, and Claude Code Sonnet~4.5 and Claude Code Haiku~4.5 trail below $1.7$.  Under multiplicity-controlled pairwise testing (Supplemental Section~\ref{supp:stats}) only the extremes of the ranking are statistically distinguishable.  Because each foundation model runs inside its own commercial harness, rank differences conflate model capability with harness design and we cannot cleanly rank the underlying models from this leaderboard alone. The more informative question is where performance breaks down, which we consider next.

\subsection{A measurement\bdash analysis gap}\label{sec:meas_gap}

\begin{figure*}
    \centering
    \includegraphics[width=0.7\columnwidth]{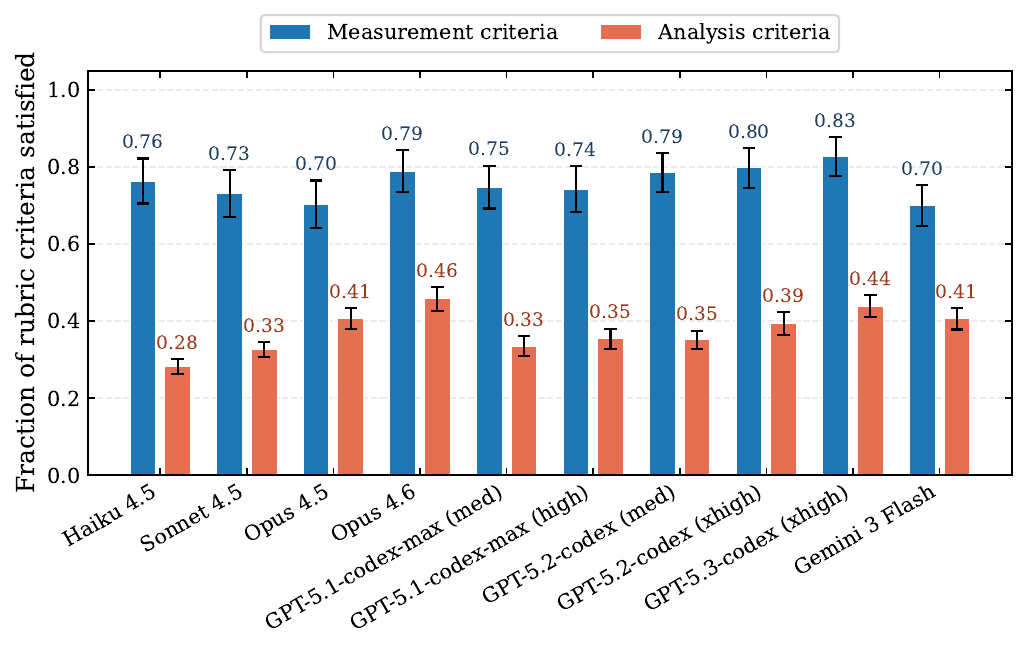}
    \caption{Mean fraction of measurement versus analysis rubric
    criteria satisfied, by agent configuration.  Error bars are
    task-clustered standard errors of the mean. }
    \label{fig:meas_analysis}
\end{figure*}

To locate where agents fail, we first split rubric criteria into measurement criteria and analysis criteria.  Measurement criteria cover instrument discovery, sweep setup, acquisition, code execution, and basic plotting.  Analysis criteria cover quantitative extraction, model/formula selection, numerical comparison, and physics reasoning. The split is stable under both an LLM classifier and a hand-curated keyword classifier (Supplemental Section~\ref{supp:meas_analysis_failures}). We are then able to split the fraction of satisfied measurement versus analysis rubric criteria, which is shown in Figure~\ref{fig:meas_analysis}.

The decomposition reveals a clear measurement\bdash analysis dissociation.  Every configuration satisfies 70--83\% of measurement criteria, whereas analysis performance ranges from 28\% for Haiku~4.5 to 46\% for Opus~4.6.  We note that although the analysis SEMs are smaller in absolute units, the SEM-to-mean ratios are comparable across the two groups (median 7.4\% measurement, 6.8\% analysis), so relative uncertainty does not differ between categories. A failure-mode classification (Supplemental Figure~\ref{fig:supp_failure_modes}) of the grade reports confirms this pattern: quantitative-analysis errors account for about 36\% of missing rubric elements, while instrument-operation errors account for only about 1.5\%.  These results indicate that, within the virtual-instrument benchmark, observed failures are dominated by scientific analysis. 

\subsection{An analysis-only ablation study}\label{sec:analysis_only}

The measurement\bdash analysis gap could reflect two different burdens.  Agents may have a residual limitation in the physics reasoning needed to analyze a dataset, or their analysis may degrade because the full measurement session introduces long context, tool-use overhead, measurement-planning decisions, and possible error propagation from earlier acquisition steps.

To separate these effects, we ran an analysis-only ablation on all 37~\qdevBench\ tasks, covering nine of the ten configurations (Gemini~3~Flash~Preview was omitted owing to model availability).  In this condition, agents received the reference CSV files directly and the prompt was rewritten to remove measurement acquisition.  No MCP server or virtual instruments were available.  The ablation setting removes the measurement session and also replaces agent-acquired data with a clean reference dataset.

\begin{figure}[!htbp]
    \centering
    \includegraphics[width=0.7\columnwidth]{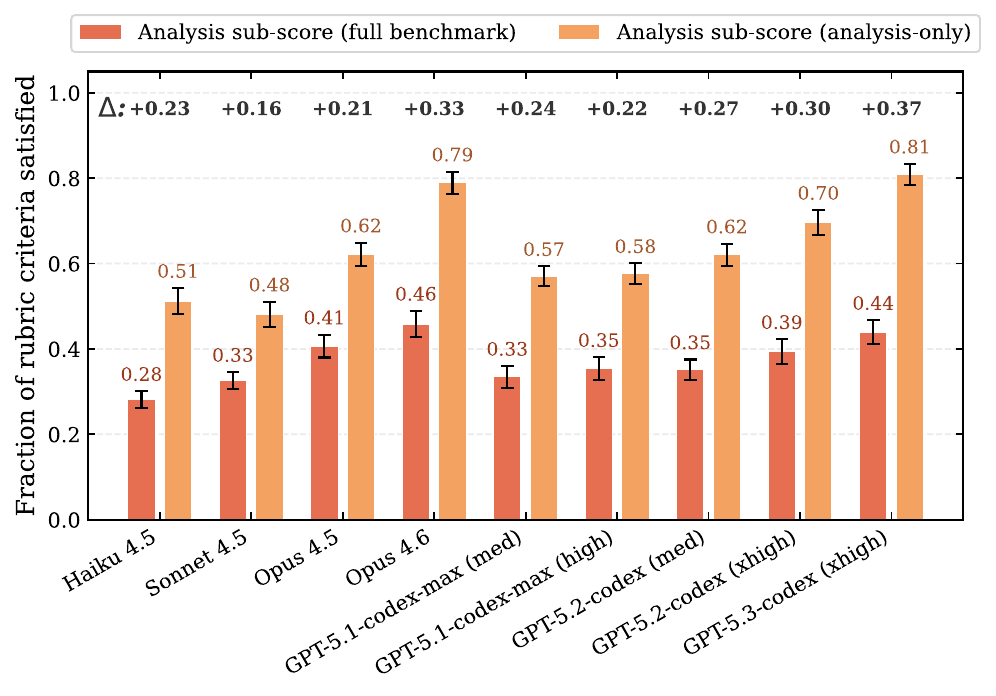}
    \caption{Analysis-only ablation for the nine configurations run in
    both conditions.  Orange-red bars show the analysis sub-score from
    full-benchmark runs, sandy-orange bars show the analysis-only
    score, and printed labels give the paired shift
    $\Delta$ (analysis-only $-$ full).  Error bars are task-clustered
    SEMs defined as in Figure~\ref{fig:meas_analysis}.}
    \label{fig:analysis_ablation}
\end{figure}

Interestingly, the ablation improves analysis performance for every tested configuration (Figure~\ref{fig:analysis_ablation}).  Across $n=333$ paired task--configuration runs, the mean shift is $\Delta=+0.26$ on the 0--1 analysis-criteria scale, with a 95\% task-clustered confidence interval (CI) of $[0.24,0.28]$.  The largest shifts occur for the strongest full-benchmark configurations: Opus 4.6 improves from 0.46 to 0.79, and GPT-5.3-codex-xhigh improves from 0.44 to 0.81. Thus, the full measurement session accounts for a substantial part of the analysis gap.  However, the best analysis-only score remains below ceiling, leaving a residual scientific-analysis gap even when clean data are provided directly.

\subsection{Scaling the benchmark}\label{sec:scaling}

Finally, we tested whether the task-construction pipeline can scale beyond the 37 expert-curated tasks.  Starting from human-selected condensed-matter papers and associated datasets, the automated pipeline generated a separate set of 47 semi-synthetic tasks, which were then post-reviewed by a domain expert and run under the full-benchmark protocol.

\begin{figure}[!htbp]
    \centering
    \begin{subfigure}[t]{\columnwidth}
        \includegraphics[width=\linewidth]{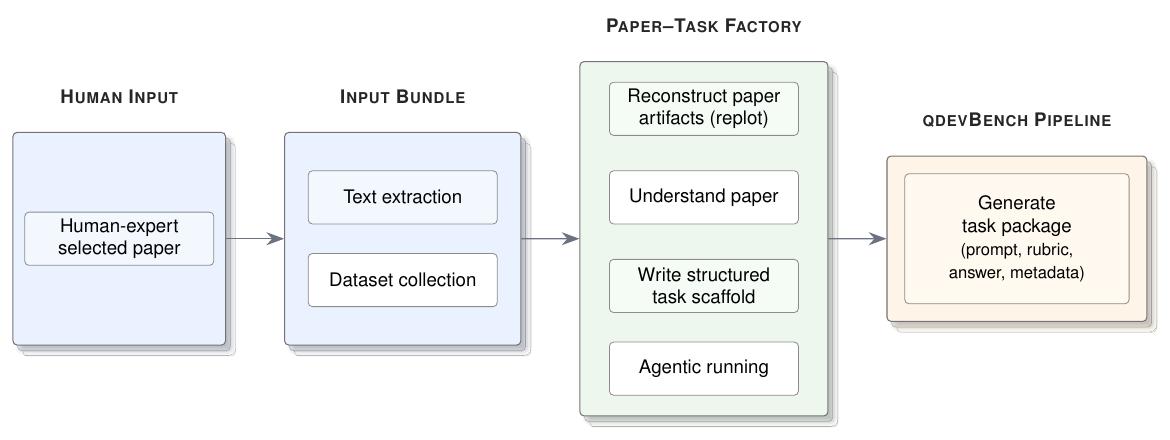}
        \caption{}\label{fig:scaleup_pipeline}
    \end{subfigure}\\[0.4em]
    \begin{subfigure}[t]{\columnwidth}
        \includegraphics[width=\linewidth]{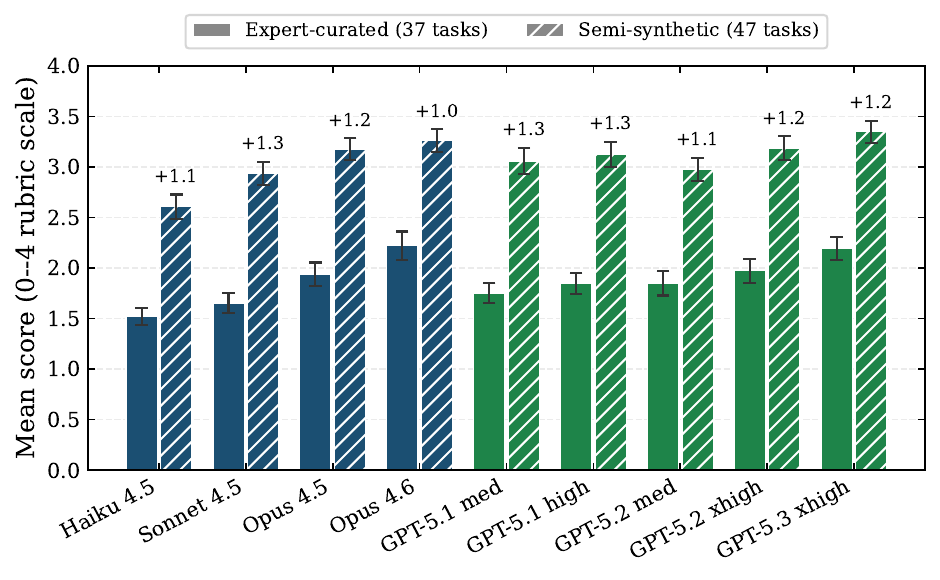}
        \caption{}\label{fig:scaleup_bars}
    \end{subfigure}
    \caption{Scaling \qdevBench\ via automated task synthesis.
    \textbf{a},~Pipeline for converting human-selected
    condensed-matter papers and associated datasets into reviewed
    \qdevBench\ task packages.  \textbf{b},~Per-configuration mean
    score on the $0$--$4$ rubric for the expert-curated set (solid
    bars, $N=37$) and semi-synthetic set (hatched bars, $N=47$), for
    the nine shared Claude~Code and Codex~CLI configurations.  Error
    bars are task-clustered $\pm 1\sigma_{\text{SEM}}$; printed labels
    give the semi-synthetic minus expert-curated shift.}
    \label{fig:scaleup_combined}
\end{figure}

The agents perform better in the semi-synthetic tasks than the expert-curated tasks (Figure~\ref{fig:scaleup_combined}).  Across the nine shared configurations, the mean score rises from $1.88$ on the expert-curated set to $3.07$ on the semi-synthetic set, a mean shift of $+1.19$ rubric points.  We therefore treat this semi-synthetic set as evidence that task supply can be scaled, but not as a direct extension of the expert-curated tasks.  The higher score distribution likely reflects narrower analytical scope and fewer multi-step domain judgments in the automatically generated tasks.  Nevertheless, the Claude~Code model hierarchy is preserved, suggesting that automated curation can expand benchmark coverage without eliminating comparative signal, which is an important property for self-evolving at scale.

\section{Discussion}\label{sec:discussion}

This study shows that current coding-agent systems can execute many
parts of a laboratory workflow, but that scientific analysis
remains the central bottleneck.  In \qdevBench, agents typically discover
instruments and run measurements, yet they often fail
to extract the correct physical quantity or apply the right domain
model.  The analysis-only ablation suggests that this failure is not
purely an intrinsic reasoning limitation. Replacing the full measurement
session with direct access to the complete reference data closes a large
fraction of the gap. At the same time, the ablation does not reach the
ceiling, indicating that residual scientific reasoning remains a real
limitation.

These results support near-term deployment of coding agents as
measurement copilots under human interpretive oversight. The safety and consent features of \instrMCP\ are designed to support this mode of use. Looking ahead, \qdevBench\ provides a controlled testbed for interventions that
target both model capability and harness design, including improved
context management, retrieval-augmented memory, and multi-agent
division of labor.

\paragraph{Limitations and future work.}
\begin{enumerate}
    \item Virtual instruments abstract away realistic experimental
    effects such as noise floors, drift, cross-talk, and hysteresis.
    Although selected tasks include controlled noise, the sandbox still
    omits much of the metrological judgment required on live hardware,
    including sweep density, dwell time, averaging, and waiting
    protocols. The sandbox also abstracts away the wall-clock and
    operational cost of each action, so agents are not rewarded for
    choosing information-dense measurements (e.g.\ a fast hysteresis
    sweep over a slow field sweep).  This tradeoff would become
    measurable only with timed virtual instruments or live deployment.
    \item High-risk hardware operations such as cryostat temperature
    ramps and cooldown cycles are out of scope here.  Extending
    \instrMCP\ to cover them would require safety interlocks beyond the
    consent framework of Section~\ref{sec:safety}.
    \item Our grading pipeline is LLM-based.  Although a stratified
    second-judge audit of the rubric stage gives
    substantial-to-almost-perfect agreement (Supplemental
    Section~\ref{supp:judge_robustness}), the Stage-1 code-analyzer
    was not independently audited.  A broad human-audited agreement
    study remains future work.
    \item The reported ranking is a model-plus-platform comparison,
    since each foundation model runs inside its native commercial
    harness.  A controlled single-harness replication is left to future
    work.
\end{enumerate}

\section*{Data and code availability}
The benchmark data, task specifications, and virtual instrument definitions are available at [URL upon publication].  The release includes the full task prompts, rubric JSON files, reference answers where applicable, and raw CSV measurement files for every task. We also provide per-run grade reports, and the extracted summary artifacts that can be used to reproduce the data in this paper.

The \instrMCP\ framework and \qdevBench\ evaluation pipeline are open-source and available at \url{https://github.com/caidish/instrMCP} and \url{https://github.com/caidish/qdevbench}.

\section*{Acknowledgements}
J.C. and P.J.-H. thank Kevin Weil for helpful discussion and support from OpenAI for Science. J.C. thanks Joshua Wang for his contributions during the early development of \instrMCP. L.-Q.X. acknowledges support from the Mathworks Fellowship. Work in the P.J.-H. lab was supported by the MIT Generative AI Impact Consortium, the MIT/Microsystems Technology Laboratories Samsung Semiconductor Research Fund, the MIT-Portugal Program, the Gordon and Betty Moore Foundation's EPiQS Initiative through Grant GBMF9463 to P.J.-H., the Heising-Simons Foundation, and by the Ramon Areces Foundation. The work at U. Washington was supported by DoE BES under the award DE-SC0018171 and the Vannevar Bush Faculty Fellowship (Award number N000142512047). I.Z. acknowledges the support from the Gordon and Betty Moore Foundation, grant DOI 10.37807/GBMF12956.

\section*{Author contributions}
J.C. and P.J.-H. conceived and designed the project. J.C. developed the \instrMCP\ framework and \qdevBench\ benchmark; S.C. developed the automated pipeline. J.C., S.C., W.S.H., L.-Q.X., S.S., S.Y., Y.Z., N.Z., W.L., and M.X. contributed to the q-devBench tasks, with input and support from I.Z., X.X., and P.J.-H. J.C., S.C., W.S.H., and L.-Q.X. wrote the manuscript with input from all authors.

\section*{Competing interests}
The authors declare no competing interests.

\clearpage
\begin{appendices}
\section{Instrument-interface implementation}\label{app:instruments}

\subsection{Safety implementation details}\label{sec:safety_impl}

\begin{figure}
    \centering
    \includegraphics[width=0.5\linewidth]{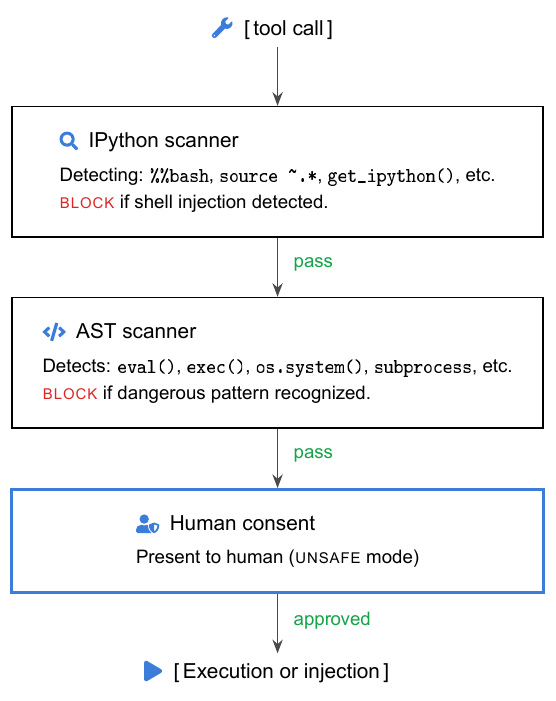}
    \caption{Safety flow for code creation and execution in a notebook.}
    \label{fig:safetyflow}
\end{figure}

\paragraph{Three-tier privilege model.}
\instrMCP\ implements a three-tier privilege escalation model with
explicit consent boundaries.  In safe mode, the agent is restricted to read-only operations such as instrument state queries, database
reads, and notebook inspection.
Unsafe mode additionally exposes notebook mutation tools (cell
execution, insertion, patching, and deletion), through which
parameter writes and sweep starts are issued as executed code, but each
invocation is intercepted by a consent dialog in the JupyterLab
frontend that the experimentalist must explicitly approve before the
operation proceeds.  Dangerous mode auto-approves all consent
requests and is intended for fully autonomous operation in
well-characterized parameter regimes or in sandboxed simulation
environments such as \qdevBench.

\paragraph{Code-scanning layer.}
Independent of the consent tier, every piece of code submitted
through the notebook mutation tools is
screened by a multi-layer static analyzer before execution
(Figure~\ref{fig:safetyflow}).  A pre-parser detects IPython magics and
shell escapes (\texttt{\%}, \texttt{!}, \texttt{get\_ipython()}),
while an abstract syntax tree~(AST) scanner identifies dangerous
Python patterns including arbitrary code evaluation (\texttt{exec},
\texttt{eval}), process spawning (\texttt{subprocess},
\texttt{os.system}), and environment manipulation.  We treat
this scanner as an additional defensive layer. This pattern-based mitigation has not been adversarially tested, and a determined caller
could in principle evade it through indirect imports,
dynamically constructed strings, or other obfuscation.  Harder
isolation in this paper is provided by operating-system- and container-level
sandboxing (Appendix~\ref{app:benchmark}, Section~\ref{sec:agents}).

\subsection{Dynamic tool subsystem}\label{sec:dynamic_impl}

The diversity of experimental protocols in table-top experiments
exceeds what any fixed tool library can anticipate, so \instrMCP\
provides a dynamic tool subsystem in which new tools can be
defined, registered, and invoked at runtime without restarting the
server.  A dynamic tool is specified by three components: a
Python function implementing its behavior, a JavaScript Object Notation (JSON) schema describing
its parameters, and a natural-language description that the agent sees
alongside the built-in tool catalog.  Upon registration, the source
code is compiled and bound to the Jupyter kernel namespace, so the
tool has direct access to all imported libraries and live instrument
handles.  The registry is persisted to disk and reloaded on server
restart, allowing experiment-specific tooling to accumulate across
sessions. Dynamic tools inherit the
same three-tier consent model and code-scanning layer as built-in
tools and are also subject to the mitigations and consent gates.  In benchmark mode each run starts from a fresh
\instrMCP\ server and a fresh Jupyter kernel with port cleanup
between runs, and the persisted dynamic-tool registry is reset to
empty before each task, so custom tools registered during one task
cannot leak into another.

\subsection{Context budget implementation}\label{sec:budget_impl}

\begin{table}[h]
\centering
\caption{Comparison of concise vs.\ detailed tool responses.}
\label{tab:concise-detailed}
\small
\setlength{\tabcolsep}{5pt}
\begin{tabular}{lll}
\toprule
\textbf{Tool} & \textbf{Concise mode} & \textbf{Detailed mode additions} \\
\midrule
\texttt{qcodes\_instrument\_info} & name, parameter list, counts & full hierarchy, validators, metadata \\
\midrule
\texttt{qcodes\_get\_parameter\_values} & instrument, parameter, value & timestamps, source, cache status \\
\midrule
\texttt{notebook\_read\_content} & source, output summary & full outputs, execution metadata \\
\midrule
\texttt{database\_list\_experiments} & \makecell[l]{experiment names, run ID ranges,\\ sweep-group summaries} & \makecell[l]{per-experiment timestamps,\\ IDs, sample names} \\
\bottomrule
\end{tabular}
\end{table}

The \instrMCP\ server stores all metadata in YAML configuration files with two layers: a
baseline file bundled with the package and an optional user file for
workflow-specific customization.  Nearly every query tool accepts a \texttt{detailed}
Boolean parameter (default \texttt{false}) that controls response verbosity
(Table~\ref{tab:concise-detailed}).  The full response is computed first, then
passed through a concise filter when \texttt{detailed=false}, ensuring that
agents can always re-query with \texttt{detailed=true} if needed.  Across the
query tools shown in Table~\ref{tab:concise-detailed}, concise mode
omits non-essential fields, freeing context budget in multi-turn
conversations.

\subsection{Agentic conversation mechanisms}\label{sec:agentic_conv_impl}

\begin{figure*}
    \centering
    \includegraphics[width=\textwidth,height=0.88\textheight,keepaspectratio]{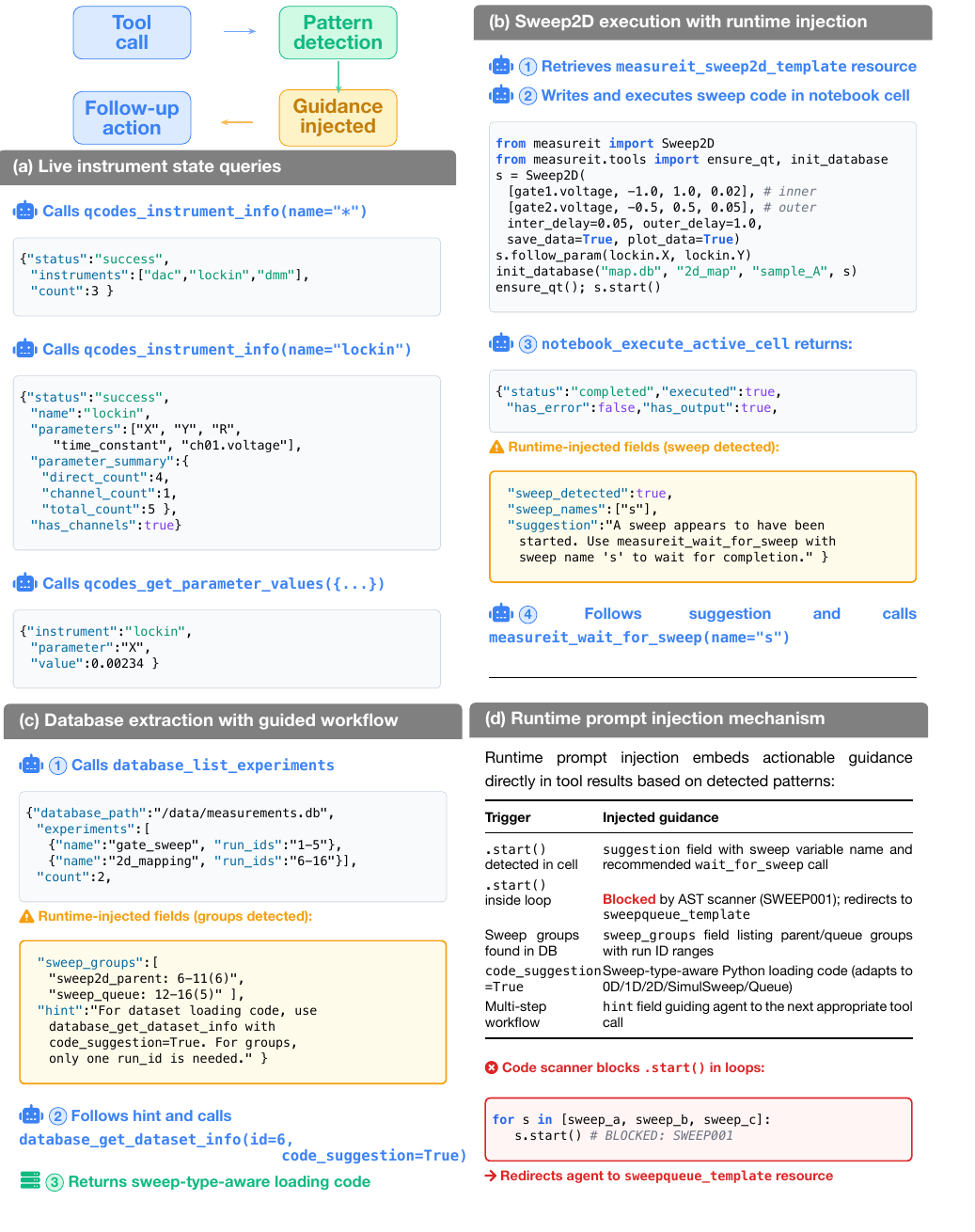}
    \caption{Agentic conversation mechanisms in \instrMCP.
    Panels summarize concise state queries, sweep templates with
    runtime guidance, database extraction hints, and the prompt-injection
    trigger flow.  The legend shows the common cycle: tool call, pattern
    detection, injected guidance, and follow-up action.}
    \label{fig:agentic_conversation}
\end{figure*}

\paragraph{Live state via tools.}
The instrument and experiment state is accessed through dedicated query tools
(Figure~\ref{fig:agentic_conversation}a).  The
\texttt{qcodes\_instrument\_info} tool returns hierarchical representations of
connected hardware, including parameter names, units, and settable ranges.
The \texttt{qcodes\_get\_parameter\_values} tool retrieves current values,
supporting both single-parameter and batch queries.  Database tools expose
recent measurements for comparative analysis.

\paragraph{Templates.}
Domain-specific resources provide working code patterns that the agent
can retrieve and adapt.  Resources such as
\texttt{measureit\_sweep1d\_template} and
\texttt{database\_access2d\_template} provide idiomatic patterns for parameter
sweeps and data retrieval (Figure~\ref{fig:agentic_conversation}b,~c).

\paragraph{Runtime prompt injection.}
\instrMCP\ employs runtime prompt injection to embed actionable guidance
directly in tool results based on detected patterns
(Figure~\ref{fig:agentic_conversation}d):
\begin{itemize}
    \item \textbf{Sweep detection.} When
    \texttt{notebook\_execute\_active\_cell} executes code containing a
    \texttt{.start()} invocation, a regex pattern detects the sweep variable
    name and injects a suggestion to call the sweep-wait tool
    (Figure~\ref{fig:agentic_conversation}b).
    \item \textbf{Database workflow hints.} The
    \texttt{database\_list\_experiments} tool detects sweep group structures
    and appends both a \texttt{sweep\_groups} summary and a \texttt{hint}
    field guiding the agent toward \texttt{database\_get\_dataset\_info}
    (Figure~\ref{fig:agentic_conversation}c).
    \item \textbf{Sweep-type-aware code generation.} The server analyzes the
    dataset's sweep type and generates appropriate loading code, automatically
    combining multi-run sweep groups into a single DataFrame
    (Figure~\ref{fig:agentic_conversation}c).
\end{itemize}

\section{Benchmark construction and agent execution}\label{app:benchmark}

\subsection{Benchmark design}\label{sec:benchmark_design}

\begin{figure}
    \centering
    \includegraphics[width=0.8\linewidth]{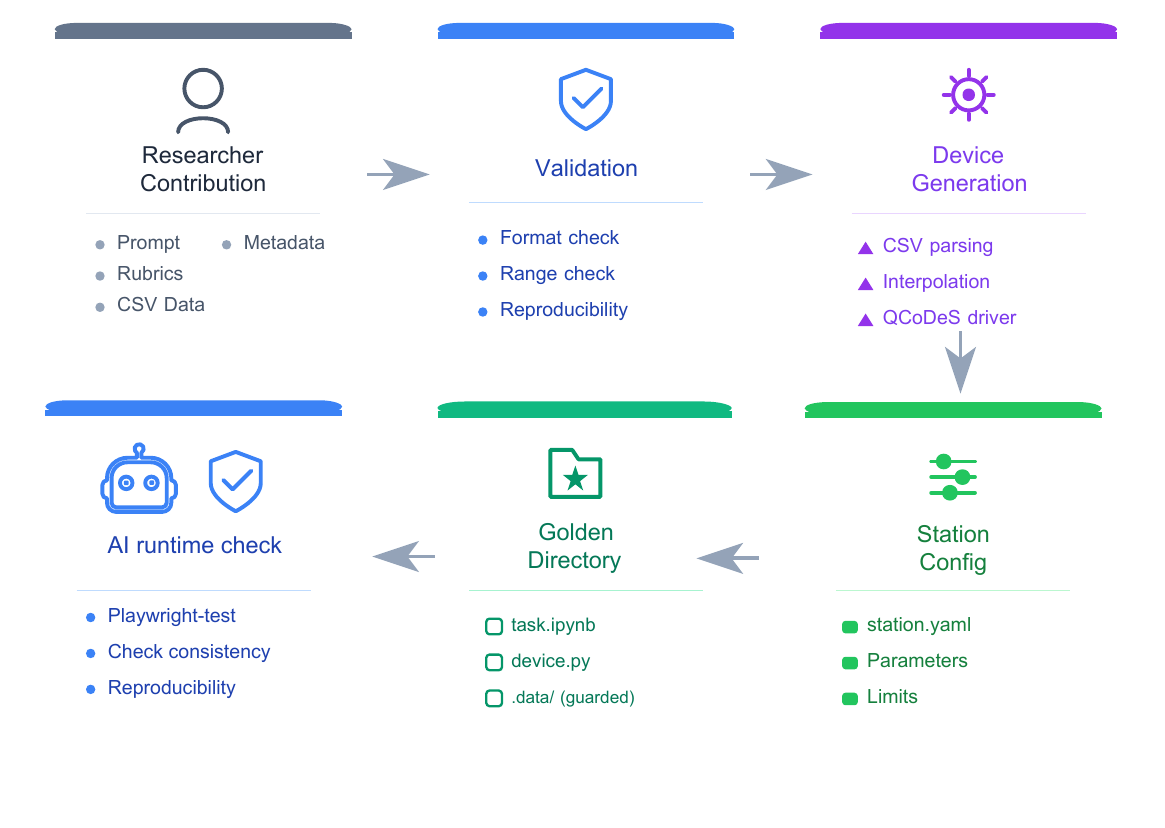}
    \caption{Data curation pipeline.  A domain expert contributes a task
    prompt, evaluation rubrics, raw CSV measurement data, and metadata.  These
    are validated, then transformed into a virtual QCoDeS instrument via
    interpolation, configured with a station file, and packaged into a
    self-contained expert-curated directory (a Golden Directory).
    Automated runtime checks verify that each packaged task is internally
    consistent and can be executed end-to-end through the benchmark
    interface before inclusion in the benchmark.}
    \label{fig:data_curation}
\end{figure}

A central challenge in benchmarking AI agents for scientific tasks
is ensuring that evaluation cannot be shortcut through memorization
of training data.  We address this through a data curation workflow
(Figure~\ref{fig:data_curation}) that builds benchmark tasks from
contributed experimental data, a mix of private measurements from
the contributing groups and published datasets, selected to be
unlikely to appear in model training corpora. 

\paragraph{Contribution structure.}
The contribution schema referenced in Section~\ref{sec:overview} is
specified as follows.  Each contribution comprises (1)~a
natural-language prompt describing the experimental goal and the
physical system under investigation, (2)~evaluation criteria
specifying what constitutes a successful response, (3)~a detailed
0--4 grading rubric, generated with LLM assistance but verified by
domain experts, (4)~raw measurement data in CSV format from actual
experiments, (5)~a dataset description covering parameter meanings and physical
units, and a source identifier (citation for
public datasets, contributor's group and run for private ones);
settable parameter ranges for the virtual devices are derived
during device generation. A
reference answer can be supplied but is not required. 

\paragraph{Automated task generation.}
For automatically synthesized tasks, the inputs are the source paper
and associated datasets, including raw instrument files and spreadsheet
data used in figures.  The curation workflow is a two-stage LLM
pipeline that supports both expert-supervised and automated runs.
Stage~1, the paper-task factory, ingests the paper-data bundle and
produces a structured task blueprint that registers the paper's
context, goal, methods, results, and claims.  A candidate scientific
question is selected from this blueprint.  Stage~2, the \qdevBench\
exporter, selects one supporting dataset slice, derives a scope
contract specifying what the selected data can and cannot support, and
emits the final task package: prompt, reference answer, evaluation
criteria, rubric, and contribution metadata.  Separating paper
comprehension from question selection and task construction prevents
the prompt from overclaiming beyond the evidence slice.  The
37~expert-curated tasks of Section~\ref{sec:results} are packaged and
validated with a domain expert supervising every stage; the
semi-synthetic task set of Section~\ref{sec:scaling} is generated
without expert intervention during construction, then post-reviewed by
a domain expert before inclusion.

\paragraph{Virtual instrument generation.}
From each contribution's measurement data, we generate a virtual QCoDeS
instrument that reproduces the parameter\bdash measurement relationships
observed in the original experimental data within the limits of the interpolation
scheme (a formal fidelity audit is reported in
Section~\ref{supp:instrument_fidelity}).  The CSV data are parsed to identify
independent parameters (setpoints) and dependent measurements (readouts). Then interpolation functions are constructed to
map parameter values to measurement outcomes. The interpolator is wrapped
in a QCoDeS-compatible driver with appropriate parameter validators, units,
and metadata.  This process is performed in collaboration with an LLM through
a curation application that automates boilerplate code generation.  The
resulting virtual instrument exposes the MCP and \QCoDeS\
interface as a real instrument, from which agents can set parameters, read
measurements, and execute sweeps, with all returned values
computed from the interpolated experimental data.

\paragraph{Station configuration.}
Each benchmark task (expert-curated and automated) includes a \texttt{station.yaml} configuration file
specifying the virtual instrument setup including the device class to instantiate, the
path to the underlying data file, initial parameter values, and settable
limits.  This mirrors the station files used in real laboratory deployments of
\instrMCP, keeping the benchmark aligned with laboratory practice.

\paragraph{Task verification.}
Before inclusion in the benchmark, each task undergoes automated
verification. The task notebook is executed to confirm that virtual instruments initialize
correctly and measurement operations complete without errors.  In
addition, an internal agent run is used as an end-to-end smoke test
of the MCP interface and task packaging.  These checks are meant to
catch broken tasks and interface regressions.

\paragraph{Task packaging.}
The final benchmark artifact is a self-contained task directory containing: (1)~a
Jupyter notebook (\texttt{task.ipynb}) with initialization cells that load the
virtual instruments and start the MCP server, while the task prompt itself is
delivered to the agent at run start through the agent CLI or an
\texttt{instruction.md} file; (2)~the \texttt{station.yaml} configuration and virtual device
source code; (3)~raw data files covered by the best-effort data-access guard; and (4)~a
hidden task specification (\texttt{.data/task.json}) containing the evaluation
rubric and any reference answer.  

\subsection{Coding agents}\label{sec:agents}

\begin{figure}
    \centering
    \includegraphics[width=1\linewidth]{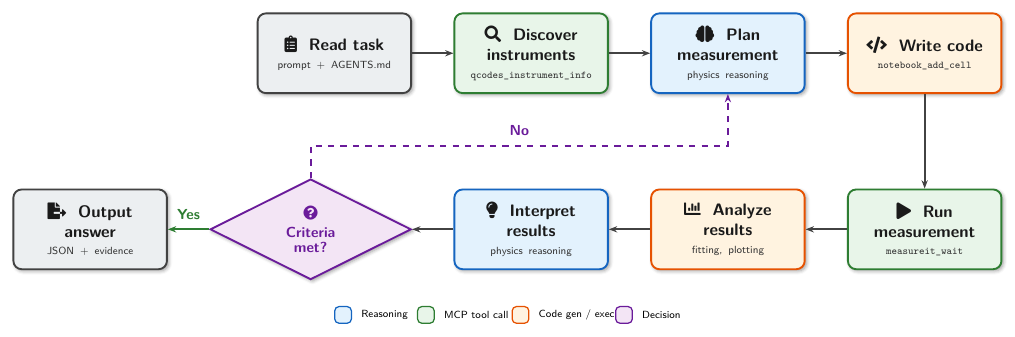}
    \caption{Typical operational loop of a coding agent during a benchmark
    task.  The agent discovers available instruments, plans measurements based
    on physics reasoning, writes and executes code in the Jupyter notebook,
    analyzes results, interprets the output, and iterates until the evaluation
    criteria are met.  Colors indicate the action type: setup/output steps
    are grey, reasoning steps (planning, interpretation) are blue,
    instrument access through MCP tool calls (including the
    instrument-driven \emph{Execute} step) is green, agent-written
    code in the notebook (writing code, running fits/plots) is
    orange, and the stopping-criterion decision is purple.}
    \label{fig:agent_loop}
\end{figure}

A typical agent loop is shown in Figure~\ref{fig:agent_loop}, and a
representative MCP tool-call sequence is shown in
Listing~\ref{lst:mcp_interaction}.  Agents
interact with the virtual instruments through MCP tool calls during
benchmark runs. Direct notebook reads of the raw experimental files
are blocked by the evaluation harness described below, so the
benchmark tests experimental reasoning rather than raw-file parsing.

\paragraph{Memory and prompt injection.}
At the start of each benchmark run, a task-agnostic instruction file (e.g.,
\texttt{AGENTS.md}) is injected into the agent's context.  This file specifies
operational constraints (e.g., no shell commands, no fabricated data),
recommended workflows for instrument discovery and measurement execution, and the expected structured response format.  The injection contains no
task-specific answers or solution strategies, ensuring that agent performance
reflects genuine problem-solving ability. An example is given in Listing~\ref{lst:agent_memory} in the Supplemental Material.

\paragraph{Sandbox and data access guard.}
For automated benchmarking, agents operate in dangerous mode, which
auto-approves all consent requests.  Actions affect only virtual
instruments and isolated QCoDeS databases inside the benchmark
sandbox.

To discourage bypassing the instrument abstraction, we deploy a
\emph{best-effort data-access guard} that blocks notebook-level reads
of the raw CSV files, the hidden \texttt{.data/} directory, and the
hidden \texttt{task.json} file containing the rubric and any reference
answer.  Direct reads of these paths raise a \texttt{PermissionError}.
The guard relies on Python stack inspection: trusted virtual-device
driver code may read the backing data, whereas agent-generated
notebook code is denied.

The strongest isolation comes from our Harbor-based Docker runners:
the task's answer directory is mounted only into the \instrMCP\
sidecar container, so the agent container has no filesystem access to
the hidden task specification and can reach its content only through
MCP tools that are designed not to return it.  At the container
boundary, data accessibility is enforced by the kernel's mount
namespace rather than by any in-process check, so the bypass categories
above do not apply.

\begin{lstlisting}[
    language=Python,
    caption={Typical MCP tool interaction during a benchmark task.  The agent
    discovers instruments (\texttt{qcodes\_instrument\_info}), writes
    measurement code to a notebook cell, executes it, waits for the sweep to
    finish, and reads results from the QCoDeS database, all through the MCP
    protocol without direct file access.},
    label={lst:mcp_interaction},
    float=t,
]
# 1. Discover available instruments
info = qcodes_instrument_info("*")
# -> magnet (field: -9..9 T), lockin (X, Y, freq)

# 2. Write measurement code to notebook cell
notebook_add_cell(content="""
from measureit import Sweep1D
sweep = Sweep1D(magnet.field, -9, 9, num=200,
                inter_delay=0.01)
sweep.follow_param(lockin.X, lockin.Y)
sweep.start()
""")

# 3. Execute cell and wait for sweep completion
notebook_execute_active_cell()
measureit_wait_for_sweep(timeout=600)

# 4. Load results from QCoDeS database
ds = database_get_dataset_info(id=1)
\end{lstlisting}

\section{Evaluation implementation and run settings}\label{app:evaluation}

\subsection{Evaluation pipeline}\label{sec:evaluation}

The two-stage LLM-as-judge pipeline of Section~\ref{sec:protocol} is
implemented as follows.

\paragraph{Result extraction.}
A unified extraction pipeline parses the agent-event stream of each run, identifies all MCP tool calls,
reconstructs the Python code written across notebook cells, and
collects the agent's final structured response.  This reconstructed
artifact set is the only agent-side input to grading. The judge
also receives the task rubric.

\paragraph{Grader inputs and bias controls.}
All agent-identifying information is anonymized before grading.  The
rubric judge uses the integer 0--4 scale of
Table~\ref{tab:grading_scale} and is instructed to prioritize
demonstrated methodology and correct numerical results over verbal
explanation quality.  LLM judges show verbosity and presentation
bias~\cite{zheng2023judging,wang2023fair}, and commercial harnesses
expose internal reasoning in different formats. Neither stage is
therefore given the agent's native \textsc{thinking}/reasoning
blocks. Both stages see only the reconstructed code, the execution
timeline, the final structured response, and the rubric.

\paragraph{Judge agreement audit.}
To bound the epistemic risk introduced by a single rubric judge we
ran a stratified second-judge audit.  For each of the
10~configurations and the three task categories (one-dimensional
[1D], two-dimensional [2D], and multi-step)
we drew a stratified $\sim$11\% sample (207 runs) spanning the four
score buckets ($0$, $1$, $2$, $3\text{--}4$), and re-graded each
selected run with an alternate rubric-stage backbone
(\textsc{gpt-5.4} at high reasoning effort) sharing the Stage-1
code-analyzer output as the primary judge, so the two judges differ
only in the rubric stage.  Per-bucket agreement counts, the
confusion matrix, the alternate judge's mean leniency, and the
rank-stability check are given in Supplemental
Section~\ref{supp:judge_robustness}.

\paragraph{Feedback collection.}
Agent responses also include structured fields for critical and soft
feedback on the \instrMCP\ interface.  This feedback informed earlier
tool-description, documentation, and error-message refinements; tool
descriptions and prompts used for the reported runs were frozen
before the evaluation campaign, and grading was performed after the
campaign in a single uniform pass with a fixed judge configuration.

\subsection{Analysis-only ablation runner}\label{sec:analysis_only_runner}

To isolate analysis capability from the measurement context in
Section~\ref{sec:analysis_only}, we built a dedicated runner without
Jupyter, \instrMCP, browser automation, or MCP tools.  For each
(task, configuration) pair, the runner creates an isolated working
directory, copies the pre-collected reference CSV files into a
\texttt{.data/} subfolder, adds an analysis-only rewrite of the task
prompt, drops an agent-memory file (\texttt{CLAUDE.md} or
\texttt{AGENTS.md}) instructing the agent to work directly against the
CSVs, and invokes the native agent command-line interface (CLI) with its usual file and shell
tools.  The reference dataset is used for every configuration and
trial.  It is not the dataset any given agent acquired during its
full-benchmark run, so the paired comparison replaces agent-acquired
data with a clean reference CSV at the same time that it removes the
measurement session.

For grading we reuse the rubric-judge stage from the main pipeline and
bypass the Stage-1 code analyzer because the analysis-only runner does
not produce the MCP/notebook execution timeline that the analyzer is
designed to summarize.  The judge receives an analysis-only prompt
variant that omits experimental-operation criteria such as sweep setup
and instrument discovery, and scores only the analysis criteria,
including physics interpretation, numerical extraction, and correctness
of the final answer.
We pair the ablation against the main benchmark at the level of
individual (task, configuration) cells.  For each of the 37 tasks
and each of the 9 configurations that were run in both the full
benchmark and the analysis-only ablation, we first average the
per-trial analysis sub-scores within the cell.  The paired observation for
that cell is the difference between the ablation mean and the
full-benchmark mean, on the 0--1 fraction-of-criteria scale.

\subsection{Experimental setup}\label{sec:exp_setup}

The configurations, platforms, trial count, and dangerous-mode
operation are specified in Section~\ref{sec:protocol}. Here we
record the remaining run-time parameters.  Trials are bounded by
per-batch wall-clock timeouts (3{,}600\,s or longer in later
batches), and Claude~Code runs additionally use the platform's turn
limit (200 turns in most batches); Codex~CLI and Gemini~CLI do not
expose an explicit turn cap.  The median run duration is about
570\,s.  Isolation
uses either the custom-built Python-guarded harness or the Harbor backend
described in Appendix~\ref{app:benchmark}, Section~\ref{sec:agents}.  Grading
uses GPT-5.2 at high reasoning effort for both the Stage-1 code
analyzer and the Stage-2 rubric judge.

\subsubsection{Models and configurations}\label{sec:models}

The 10 evaluated configurations and their canonical names
(GPT-5.1-codex-max at medium and high effort, GPT-5.2-codex at
medium and extra-high (xhigh) effort, GPT-5.3-codex at xhigh effort,
Claude Haiku~4.5, Sonnet~4.5, Opus~4.5, and Opus~4.6, each at high
effort, and Gemini~3~Flash~Preview) are listed in
Supplemental Table~\ref{tab:model_results} and used consistently throughout the
paper; the selection spans model families, sizes, platforms, and
inference-time compute budgets.

\end{appendices}

\clearpage
\bibliographystyle{unsrt}
\bibliography{refs}

\clearpage
\begin{center}
  {\LARGE Supplemental Material for\\[0.35em]
   \large Evaluating Instrument-Centered AI Agents for Table-Top Science Experiments\par}
\end{center}
\vspace{1em}


\setcounter{figure}{0}
\renewcommand{\thefigure}{S\arabic{figure}}
\def\theHfigure{S\arabic{figure}}
\setcounter{table}{0}
\renewcommand{\thetable}{S\arabic{table}}
\def\theHtable{S\arabic{table}}
\setcounter{lstlisting}{0}
\def\thelstlisting{S\arabic{lstlisting}}
\def\theHlstlisting{S\arabic{lstlisting}}

\setcounter{section}{0}
\renewcommand{\thesection}{S\arabic{section}}
\renewcommand{\theHsection}{supp.\arabic{section}}

\section*{Supplemental analyses}\label{sec:supp}

This supplement is organized as six short notes.  Notes~1--3 describe
the benchmark task set, virtual-instrument fidelity and grading agreement audits that establish its credibility.  Notes~4--5
support the main-text statistical claims.  Note~6 reproduces the
task-agnostic agent instruction file.
\section{Benchmark task set and packaging}\label{supp:tasks}

\qdevBench\ consists of 37~tasks reconstructed from real
condensed-matter experiments (Appendix~\ref{app:benchmark}, Section~\ref{sec:benchmark_design}).
Each task packages a natural-language prompt, a reference answer
where a single expected answer is defined, evaluation criteria, a
detailed 0--4 rubric, raw CSV measurement data, and metadata
specifying parameter descriptions, physical units, and valid
ranges, and is frozen together with a matching \instrMCP\ version.

Table~\ref{tab:supp_task_set} lists every task.  The
\textsc{Q} column marks whether the task's headline
deliverable is a specific numerical value (e.g.\ critical
temperature, coercive field, capacitance, activation gap); tasks
whose correct answer is primarily a state identification or phase
label are left blank even if they quote a number in passing.  The
\textsc{Pub} column marks whether a public reference dataset was
used.

\begin{table}[!htbp]
\centering
\caption{The 37~\qdevBench\ tasks, grouped by task type.
\textsc{Type} classifies the measurement protocol: 1D (single
parameter sweep), 2D (two-parameter sweep or image), or
multi-step (tasks that require combining more than one
measurement, e.g.\ sweeps at several temperatures or a sweep
plus a follow-up calibration).  Multi-step is a protocol
category, not a sweep dimensionality.
\textsc{Q} = primary deliverable is a numerical value.
\textsc{Pub} = based on a publicly released dataset.
Acronyms: RMCD, reflective magnetic circular dichroism;
TDBG, twisted double bilayer graphene;
MATBG (MATTG), magic-angle twisted bilayer (trilayer) graphene;
TLG, trilayer graphene;
hBN, hexagonal boron nitride;
NV, nitrogen-vacancy;
AFM, atomic force microscopy;
AM (FM), amplitude (frequency) modulation;
SPM, scanning probe microscopy;
STM, scanning tunneling microscopy;
STP, scanning tunneling potentiometry;
LDOS, local density of states;
CDW, charge density wave;
QPI, quasiparticle interference;
BKT, Berezinskii\bdash Kosterlitz\bdash Thouless;
PZT, lead zirconate titanate;
YIG, yttrium iron garnet.}
\label{tab:supp_task_set}
\scriptsize
\begin{tabular}{@{}llllp{0.55\columnwidth}@{}}
\toprule
Task ID & Type & Q & Pub & Short title \\
\midrule
UKQ5TFJBH & 1D & Y & Y & RMCD hysteresis \& coercive field of MnBi2Te4 device \\
X9BJE6PR4 & 1D & Y & Y & Quantum Hall onset field in MnBi2Te4 \\
0I5IEDXZ8 & 2D &  & Y & Correlated insulators in TDBG via Rxx gate maps \\
0O0MNB7WX & 2D &  & Y & Landau fan of MATBG to identify Chern insulators \\
22J3OAL1N & 2D & Y &  & NV microscopy: AM vs FM-AFM sensor-sample distance \\
2IGTQSJV6 & 2D &  & Y & STM topography of CsV3Sb5-xSnx symmetry breaking \\
5ROJQE7KL & 2D & Y & Y & Quantized Hall phase boundary in MnBi2Te4 \\
66THLKHA7 & 2D & Y &  & Magnetic field sensor displacement calibration \\
6IGL3OZQ6 & 2D & Y &  & Vortex domain identification in YMnO3 surface scan \\
6R4JQ33MR & 2D & Y &  & Magnetic susceptibility from SPM field gradient map \\
7MBF4YDLO & 2D &  & Y & Tunneling spectroscopy of twisted trilayer graphene \\
853YRG42X & 2D & Y &  & BKT transition temperature from V--I curves at multiple temperatures \\
938B0XJWO & 2D & Y & Y & Landau fan of MoTe2: extract Chern numbers \\
9DX1NB73A & 2D & Y & Y & Curie temperature from Hall resistance vs B and T \\
9MLITYNUD & 2D & Y & Y & Twist angle from dual-gate resistance map of TLG \\
HTWKDCRSE & 2D &  & Y & Van Hove singularity in twisted trilayer graphene \\
K187BQTUM & 2D &  & Y & Electronic nematicity and symmetry in CsTi3Bi5 LDOS \\
KCDNTTO8I & 2D & Y & Y & Quantized Hall phases in MnBi2Te4 phase diagram \\
L6SBCPIX1 & 2D & Y &  & Stroboscopic magnetic sensor oscillation amplitudes \\
MC58JW8U9 & 2D &  & Y & CDW detection in CsTi3Bi5 via LDOS mapping \\
OUD5DKXT3 & 2D &  & Y & Landauer residual resistivity dipole in graphene STP \\
P1J2AFA5D & 2D &  &  & NV magnetometry of magnetic racetrack with domain walls \\
PP68HC41C & 2D & Y &  & AFM step-edge height analysis of Cr2O3 (0001) \\
RN0UVL4V3 & 2D &  &  & Quantum Hall bilayer phase diagram in graphene/hBN \\
RSB1KV3S5 & 2D &  & Y & Rotational symmetry breaking in kagome metal KV3Sb5 \\
RSB2SCV6S & 2D &  & Y & Rotational symmetry breaking in kagome ScV6Sn6 \\
S65QOMK9A & 2D &  &  & Interlayer tunneling spectroscopy in MATTG heterostr\ldots \\
TF7IHQN5S & 2D & Y &  & Landau fan and top-gate capacitance extraction \\
URHV8S4K6 & 2D & Y & Y & Landau fan and Chern number extraction in MoTe2 \\
WJMQRITOP & 2D & Y &  & Thermal activation gap of Chern insulator in TLG \\
Y29A4LUT3 & 2D &  &  & Domain wall charge model for PZT thin film \\
2ZSJC6LVW & multi &  & Y & Dirac QPI band splitting in TbV6Sn6 under field \\
GTZ5B1GG6 & multi &  &  & Superconductivity test via longitudinal resistance \\
JMON8LCGI & multi & Y &  & Magnon transport characterization in YIG/Pt device \\
UNX0Q8RWY & multi & Y & Y & Twisted MoTe2 quantum Hall gap extraction \\
X7XC7C1CD & multi & Y &  & RuO2 thermometer calibration and self-heating study \\
Y624HVE4B & multi &  &  & Gap nature from tunneling \& BKT in twisted trilayer \\
\bottomrule
\end{tabular}
\end{table}

\section{Virtual-instrument fidelity}\label{supp:instrument_fidelity}

The virtual instruments used in \qdevBench\ are constructed by
interpolating contributed CSV measurement data
(Appendix~\ref{app:benchmark}, Section~\ref{sec:benchmark_design}).  To quantify the on-grid
reproduction of this construction we queried every instrument
at the original experimental data points and compared the
returned values against the ground-truth CSV.  For tasks with
fewer than 5{,}000 data points all rows were used; larger
datasets (up to 2.8\,M rows) used task-dependent checks:
typically stratified random samples of about 900--1{,}800 points,
with larger or full-grid audits for selected datasets.  For the
largest datasets with sampled checks, the sampled query points,
expressed in physical coordinates, need not coincide exactly with
the reconstructed grid nodes, so the check effectively becomes an
off-node audit for those datasets.  Most virtual devices are
deterministic; for the two devices with built-in runtime noise
(Section~\ref{sec:framework}), the audit measures the noisy output.
We report normalized
root-mean-squared error (NRMSE), defined as RMSE divided by
the range of the measured quantity and expressed as a
percentage.  Because regular-grid interpolators pass through
the input points by construction, on-grid NRMSE is
mechanically zero for most tasks; this audit is therefore a
reproduction check at the sampled points, not a validation of
off-grid interpolation fidelity or of the parameter regimes
agents actually query during a run.

\begin{figure}[!htbp]
    \centering
    \includegraphics[width=\columnwidth]{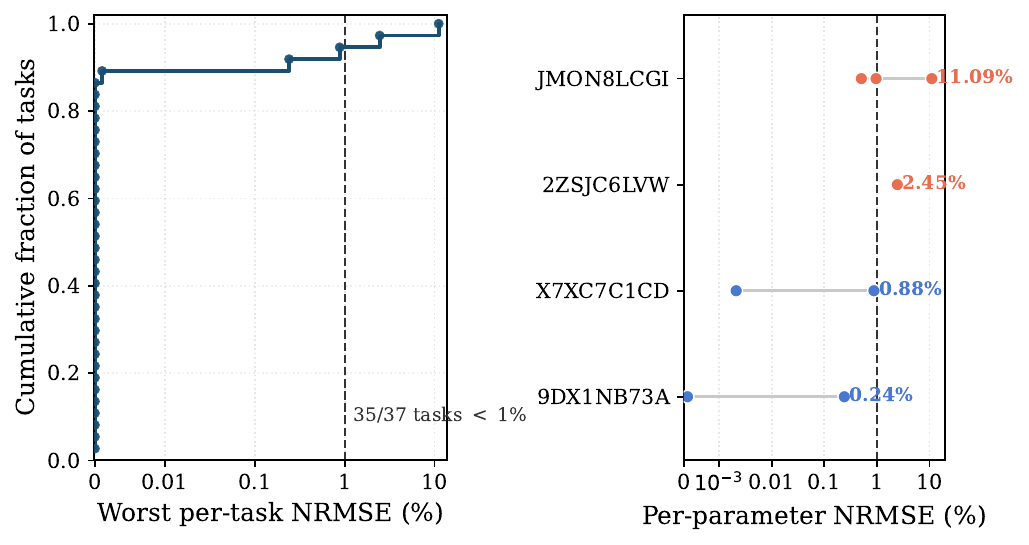}
    \caption{Normalized RMSE of virtual-instrument output versus
    original experimental data for all 37~benchmark tasks, sorted by
    error magnitude.  The dashed line marks the 1\% threshold.
    35 of 37 tasks fall below 1\%; the two outliers are a
    scattered-interpolation STM dataset (2.4\%) and a single
    physics-model device (11\% on one output channel).}
    \label{fig:instrument_fidelity}
\end{figure}

Most tasks use regular-grid interpolation
(\texttt{RegularGridInterpolator}) and achieve exactly 0\% NRMSE
when queried at the original grid coordinates by construction (see Figure~\ref{fig:instrument_fidelity}),
since the interpolator passes through the input points.  Two tasks
exceed 1\%: for QDB-2ZSJC6LVW (a 2.8\,M-row STM~$dI/dV$ map
reconstructed on a regular three-dimensional grid per magnetic
field), the audit's stratified sample of $\sim$1{,}400 query
points is expressed in physical coordinates that need not coincide
exactly with the reconstructed grid nodes, giving 2.4\% NRMSE
from linear interpolation between neighboring grid nodes.
QDB-JMON8LCGI is the single physics-model device in
the benchmark, implementing an analytic transfer function fitted to
the experimental data, and its detector-phase channel shows 11\%
NRMSE, reflecting model approximation error together with the
device's built-in measurement noise rather than an interpolation
artifact.

\section{Grading agreement}\label{supp:judge_robustness}

Because all 1{,}850~grades in this paper come from a single
two-stage LLM judge (Appendix~\ref{app:evaluation}, Section~\ref{sec:evaluation}), the judge is the
largest single source of epistemic risk in the pipeline.  To bound
this risk we ran a stratified second-judge audit: for each of the
10~configurations and each of the three task categories
(1D, 2D, multi-step) we drew a stratified
per-bucket sample (207~runs, $\sim$11\% of all grades) spanning the four score buckets
($0$, $1$, $2$, $3\text{--}4$).  Each selected run was re-graded
with an independent judge LLM (\textsc{gpt-5.4} at high
reasoning effort) sharing the Stage-1 code-analyzer output as
the primary judge, so the two judges differ only in the rubric
stage.  This audit therefore bounds only the rubric-stage
contribution to judge variability; any systematic bias injected
by the shared Stage-1 code analyzer is not detected here,
because it propagates identically into both rubric judges.  A
fully independent two-stage regrade on a subset would tighten
this bound and is left to future work.

\begin{figure}[!htbp]
    \centering
    \includegraphics[width=0.85\columnwidth]{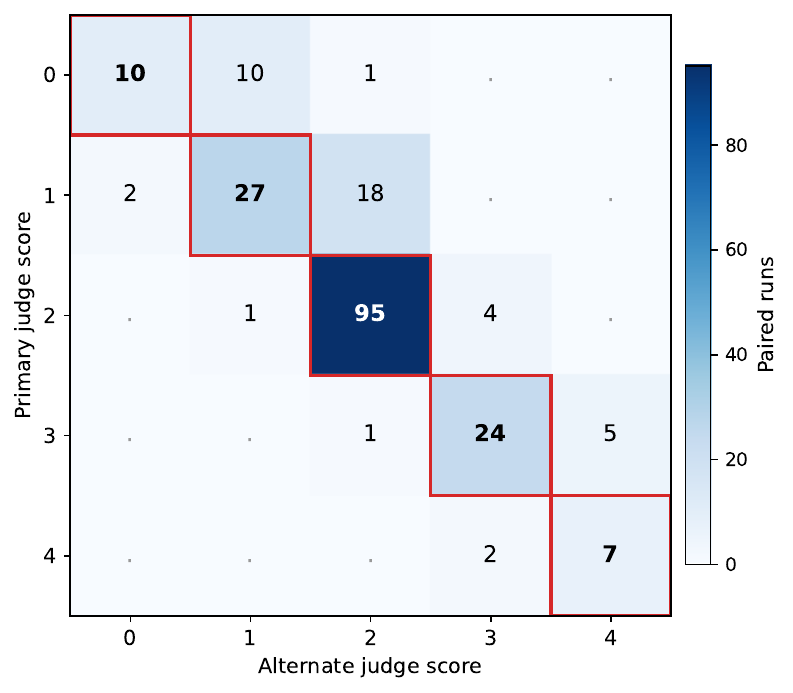}
    \caption{Second-judge agreement on the 207-run stratified audit.
    Rows: primary \textsc{gpt-5.2} judge score.  Columns: alternate
    \textsc{gpt-5.4} judge score.  Cell entries are counts of runs.}
    \label{fig:judge_robustness}
\end{figure}

Across the 207 paired grades, 163~runs (79\%) received the same
score from both judges; 37~(18\%) differed by $+1$ rubric point, six by
$-1$, and one by $+2$. Figure~\ref{fig:judge_robustness} compares the two judges' scores.  The alternate judge was modestly more
lenient, by $+0.16$ rubric points on average.  Reproducing the
ranking of Table~\ref{tab:model_results} under the alternate
judge preserved the top pair (Opus~4.6 and GPT-5.3-codex) and
the bottom two configurations (Sonnet~4.5 and Haiku~4.5), with
no rank flip larger than one position in the middle tier.  We
therefore treat the numerical values in the main text as
judge-dependent at the $\pm 0.2$-rubric level; the coarse
ranking extremes are stable, while middle-tier ordering
remains uncertain.

\section{Statistical analysis of benchmark scores}\label{supp:stats}

For each configuration we first average the 5~trial scores within a
task to obtain 37~task-level means, and base all uncertainty
statements on these 37~values: bootstrap confidence intervals are
obtained by resampling the 37~tasks with replacement
($10{,}000$~resamples; percentile method), and the error bars in
Figures~\ref{fig:forest}, \ref{fig:meas_analysis}, and
\ref{fig:analysis_ablation} are task-clustered SEMs,
$\mathrm{std}(\text{task means}) / \sqrt{n_\text{tasks}}$.

For omnibus comparison across configurations we apply a Friedman
test on the task-level paired scores
($\chi^2_F = 88.1$, $\mathrm{df} = 9$, $p < 10^{-14}$), followed by
a Nemenyi post-hoc test for multiplicity-controlled rank separation
(critical difference~$\mathrm{CD} = 2.23$ at $\alpha = 0.05$; 17 of
45~pairs reach significance).  The Friedman test rejects equal
performance across configurations, but Nemenyi separation is
concentrated between the strongest and weakest configurations; most
adjacent configurations are not separated.  The main text
therefore interprets the configuration ranking as contextual rather
than as a load-bearing scientific result
(Section~\ref{sec:overall}, Table~\ref{tab:model_results}).

\begin{table}[!htbp]
\centering
\caption{Configuration performance summary.  Mean score on the
0--4 rubric scale with task-clustered bootstrap 95\% CIs
($10{,}000$ resamples of the 37~tasks with within-task trial
averaging; percentile method).  Configurations ranked by mean.}
\label{tab:model_results}
\small
\begin{tabular}{@{}lcc@{}}
\toprule
Configuration & Mean & 95\% CI \\
\midrule
Opus 4.6                     & 2.22 & [1.95, 2.49] \\
GPT-5.3-codex (xhigh)        & 2.19 & [1.98, 2.42] \\
GPT-5.2-codex (xhigh)        & 1.97 & [1.74, 2.21] \\
Gemini 3 Flash Preview       & 1.95 & [1.71, 2.18] \\
Opus 4.5                     & 1.94 & [1.72, 2.16] \\
GPT-5.2-codex (medium)       & 1.85 & [1.62, 2.08] \\
GPT-5.1-codex-max (high)     & 1.84 & [1.64, 2.04] \\
GPT-5.1-codex-max (medium)   & 1.75 & [1.57, 1.94] \\
Sonnet 4.5                   & 1.65 & [1.47, 1.83] \\
Haiku 4.5                    & 1.52 & [1.35, 1.69] \\
\bottomrule
\end{tabular}
\end{table}

\paragraph{Paired Wilcoxon recipe.}
For all paired comparisons (analysis-only ablation in
Appendix~\ref{app:evaluation}, Section~\ref{sec:analysis_only_runner} and the
reasoning-effort pairs below) we use the same statistical recipe.
The primary significance test is a task-level Wilcoxon signed-rank
test: we average the paired cell deltas across all configurations
within each task to obtain 37 task-level paired observations, and
test the null of zero-median paired difference.  This matches the
dependence structure of the task-clustered bootstrap CI, which
resamples entire tasks with replacement rather than independent
cells, treating tasks as the random effect when generalizing
beyond the 37 tasks evaluated.  As a secondary check we also
compute the signed-rank test over the 333 individual
(task, configuration) cells.  Wilcoxon is used in preference to a paired
$t$-test because the sub-scores are bounded in $[0,1]$ and heavily
skewed at the extremes, which violates normality; the signed-rank
statistic uses only the rank and sign of the paired differences
and is less sensitive to the non-Gaussian tails.

\begin{figure}[!htbp]
    \centering
    \includegraphics[width=0.9\columnwidth]{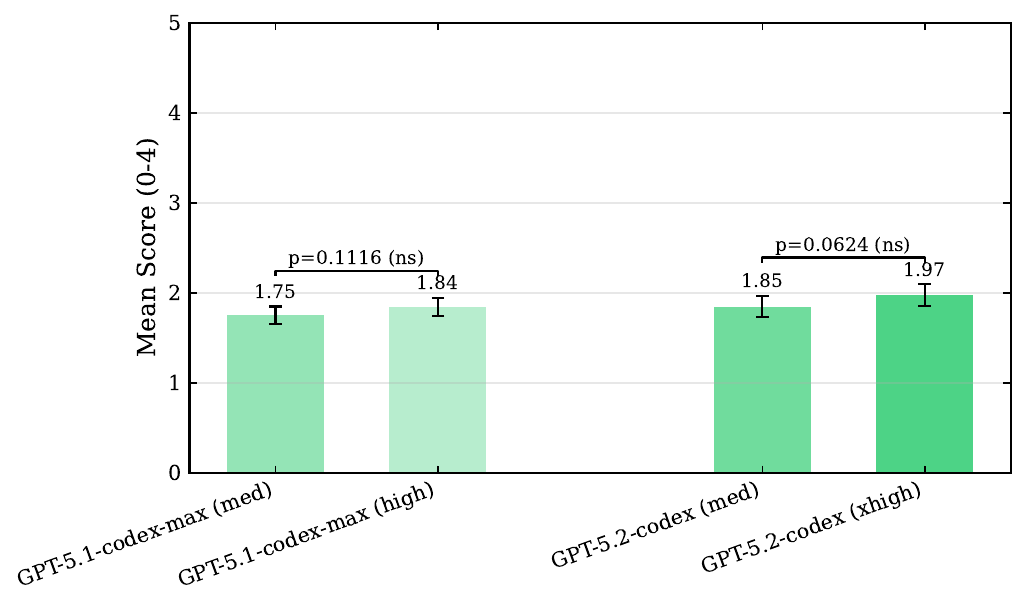}
    \caption{Paired reasoning-effort comparisons for the two
    Codex families.  Bars show the mean score on the 0--4 rubric
    for each configuration, with error bars giving task-clustered
    $\pm 1\,\sigma_{\text{SEM}}$ (per-task means as the
    independent replicate; $T=37$).  The paired shift and the
    Wilcoxon signed-rank $p$-value are printed above each pair.}
    \label{fig:supp_reasoning_summary}
\end{figure}

\paragraph{Reasoning-effort ablation.}\label{supp:reasoning}%
Within the Codex family we also compared pairs of configurations
that share the base model and differ only in reasoning effort.
Increasing effort produced small, non-significant improvements on
the 0--4 scale: $+0.09$ for GPT-5.1-codex-max
(medium$\to$high; Wilcoxon signed-rank $p = 0.112$) and $+0.12$
for GPT-5.2-codex (medium$\to$xhigh; $p = 0.062$).  As a
per-task check we also ran two-sided Mann\bdash Whitney $U$ tests
on the raw trial scores for each task and effort pair (five trials
per effort setting, without assuming matched seeds between effort
settings); only 1 of 74 per-task tests reached $p < 0.05$, within
the range expected by chance at that multiplicity.  We therefore do
not find statistically reliable evidence that additional
reasoning-effort budget closes the analysis gap at the effort
levels tested.  This is a null result with limited power
($T = 37$ paired tasks) and effect sizes near $+0.1$ rubric
points.  Figure~\ref{fig:supp_reasoning_summary} summarizes
the paired means, shifts, and signed-rank $p$-values for both
effort comparisons.

\section{Measurement\bdash analysis decomposition and failure modes}\label{supp:meas_analysis_failures}

The measurement\bdash analysis decomposition reported in
Figure~\ref{fig:meas_analysis} classifies every rubric element
string into a measurement category (instrument operation,
data acquisition, code execution, plotting) or an analysis
category (scientific interpretation, quantitative extraction,
physics reasoning) using an LLM classifier (GPT-4o-mini).  Across
all graded rubric elements, 76\% were classified as analysis and
24\% as measurement, reflecting an analysis-heavy rubric design.
As a check we re-classified the elements with a
best-effort hand-curated keyword-based classifier.  A rubric
element is tagged as measurement if (and only if) it
contains a substring match against the measurement keyword list
below and no match against the analysis keyword list, and
analogously for analysis. The full keyword lists are reproduced below.

\paragraph{Measurement keywords.}
{\small
\begin{itemize}\setlength{\itemsep}{2pt}
  \item \textit{Sweep / acquisition}: \texttt{sweep}, \texttt{measure}, \texttt{acqui}, \texttt{record}, \texttt{save}, \texttt{instrument}, \texttt{discover}, \texttt{parameter}, \texttt{data}, \texttt{channel}, \texttt{lock-in}, \texttt{lockin}, \texttt{monitor}, \texttt{configur}.
  \item \textit{Timing / delays}: \texttt{dwell}, \texttt{delay}, \texttt{inter\_delay}, \texttt{outer\_delay}, \texttt{settling}, \texttt{wait}, \texttt{timeout}, \texttt{stabiliz}.
  \item \textit{Code / execution}: \texttt{code}, \texttt{notebook}, \texttt{cell}, \texttt{execut}, \texttt{implement}, \texttt{script}, \texttt{runs}, \texttt{run}, \texttt{ran}.
  \item \textit{Plotting basics}: \texttt{plot}, \texttt{colormap}, \texttt{pcolormesh}, \texttt{imshow}, \texttt{colorbar}, \texttt{axes}, \texttt{label}, \texttt{figure}, \texttt{saved}.
  \item \textit{Hardware / instrument specifics}: \texttt{voltage}, \texttt{current}, \texttt{field}, \texttt{temperature}, \texttt{gate}, \texttt{magnet}, \texttt{resistance}, \texttt{Rxx}, \texttt{Rxy}, \texttt{Vxx}, \texttt{Vxy}, \texttt{nested}, \texttt{grid}, \texttt{loop}, \texttt{num\_points}, \texttt{num=}, \texttt{points}.
  \item \textit{Database / file}: \texttt{database}, \texttt{QCoDeS}, \texttt{db}.
\end{itemize}
}

\paragraph{Analysis keywords.}
{\small
\begin{itemize}\setlength{\itemsep}{2pt}
  \item \textit{Scientific interpretation}: \texttt{identif}, \texttt{extract}, \texttt{interpret}, \texttt{analy}, \texttt{comput}, \texttt{calcul}, \texttt{conclud}, \texttt{determin}, \texttt{attribut}, \texttt{threshold}, \texttt{interference}, \texttt{pattern}, \texttt{peak}, \texttt{critical}, \texttt{quantit}, \texttt{convert}, \texttt{formula}, \texttt{correct value}, \texttt{onset}, \texttt{calibrat}, \texttt{fit}, \texttt{phase}, \texttt{topolog}.
  \item \textit{Physics concepts}: \texttt{Chern}, \texttt{filling}, \texttt{Landau}, \texttt{Josephson}, \texttt{superconduct}, \texttt{ferromagnet}, \texttt{insulating}, \texttt{Van Hove}, \texttt{gap}, \texttt{BKT}, \texttt{transition}, \texttt{symmetry}, \texttt{correlated}, \texttt{order}, \texttt{invariance}, \texttt{wavelength}, \texttt{harmonic}, \texttt{spectral}, \texttt{conductance quantiz}, \texttt{plateau}, \texttt{quantum Hall}, \texttt{activation}, \texttt{energy}, \texttt{dispersion}.
  \item \textit{Numerical results}: \texttt{within tolerance}, \texttt{ground.truth}, \texttt{reported value}, \texttt{numeric}, \texttt{value}, \texttt{result}, \texttt{answer}, \texttt{correctly}, \texttt{accurately}, \texttt{match}.
  \item \textit{Domain wall / imaging analysis}: \texttt{domain}, \texttt{vortex}, \texttt{defect}, \texttt{dislocat}, \texttt{cross.correlat}, \texttt{displacement}, \texttt{sensor}.
  \item \textit{Math / signal processing}: \texttt{FFT}, \texttt{fft}, \texttt{Fourier}, \texttt{periodicity}, \texttt{decay}, \texttt{model}, \texttt{exponential}, \texttt{spatial}, \texttt{frequency}, \texttt{anisotrop}, \texttt{background subtract}, \texttt{background suppress}.
  \item \textit{Inference / reasoning}: \texttt{inference}, \texttt{proximity}, \texttt{closer}, \texttt{larger signal}, \texttt{hypothesis}, \texttt{evidence}, \texttt{conclude}, \texttt{explanation}, \texttt{consistent with}, \texttt{indicates}.
  \item \textit{Comparison / evaluation}: \texttt{comparison}, \texttt{compare}, \texttt{differ}, \texttt{agree}, \texttt{discrepan}, \texttt{error}, \texttt{deviation}, \texttt{offset}.
  \item \textit{Visualization of analysis}: \texttt{Wannier}, \texttt{fan}, \texttt{$\rho_{xx}$}, \texttt{$\rho_{xy}$}, \texttt{rho}, \texttt{color map}, \texttt{heatmap}.
\end{itemize}
}

\noindent Applying this classifier to the rubric elements,
the per-configuration measurement and analysis fractions change
by at most 1--2 percentage points relative to the primary
LLM-based classification, and the measurement\bdash analysis gap
is unchanged.  We also confirmed that error bars in
Figures~\ref{fig:meas_analysis} and \ref{fig:analysis_ablation}
are task-clustered SEMs: per-run fractions are first averaged
within each task and SEM is then taken across the 37~task means, so
uncertainty does not artificially shrink with the number of seeds.

To locate where agents fail we classified every missing-element
and common-mistake entry in the 1{,}850~grade reports into seven
failure categories using a keyword-based classifier.  Because the
classifier is heuristic, we use the resulting counts only as a
coarse diagnostic and not as a precise causal decomposition.

\begin{figure}[!htbp]
    \centering
    \includegraphics[width=\columnwidth]{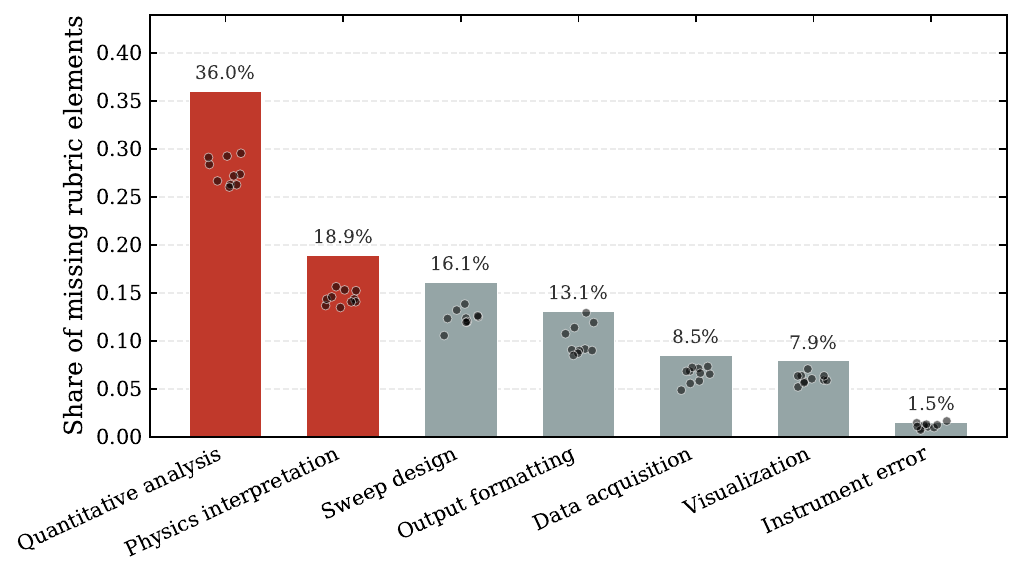}
    \caption{Share of missing rubric elements by failure category,
    aggregated across all 10~configurations.  Bars: aggregate share.
    Dots: per-configuration share (one dot per configuration,
    slight horizontal jitter).  Analysis-side categories
    (quantitative analysis, physics interpretation) are highlighted
    in red and together account for more than half of the missing
    rubric elements; instrument-operation errors are rare.  The
    catch-all other bucket is omitted for readability.}
    \label{fig:supp_failure_modes}
\end{figure}

Two features of Figure~\ref{fig:supp_failure_modes} are stable under
the heuristic nature of the classifier.  First, the two analysis
categories (quantitative analysis and physics interpretation)
together dominate the failure budget, which is consistent with the
measurement\bdash analysis gap identified in the main text.
Second, configurations have broadly similar per-category shares, indicating that the failure mix is driven by the benchmark. Configurations differ mainly in failure rate, not failure type.
Instrument-operation errors account for about 1.5\% of the missing
rubric budget, consistent with the near-uniform measurement
competence in the main text.

\section{Task-agnostic agent instruction file}\label{supp:agent_memory}

The task-agnostic instruction file is injected into every run
(Section~\ref{sec:framework}), written as \texttt{AGENTS.md},
\texttt{CLAUDE.md}, or \texttt{GEMINI.md} depending on the agent.  It is
identical across all 37~tasks and all configurations, and contains no
task-specific answers or solution strategies.  Listing~\ref{lst:agent_memory}
reproduces it verbatim.

\begin{lstlisting}[
    language={},
    caption={The task-agnostic instruction file injected at the start of
    every benchmark run, reproduced verbatim.  It specifies operational
    constraints, a recommended discovery\bdash measurement\bdash analysis
    workflow, and the structured JSON response format.  It is identical
    across all tasks and configurations.},
    label={lst:agent_memory},
]
# Instructions
Use **MCP tools** to interact with lab instruments to perform experiments and collect data, and use the **Jupyter notebook** to accomplish condensed matter physics research tasks. All work will be verifiable by a human researcher inside the notebook. Plot things nicely for them to exam your work. Be concise and precise.

## Hard Constraints (Do Not Violate)
- **No shell usage**: do not run shell commands, source rc files, set environment variables, or activate conda environments.
- **No fabricated data**: only report results produced by executed notebook code and completed MeasureIt sweeps.
- **Default to `detailed=false`** for tool calls unless details are necessary for diagnosis.
- **Tool/sweep failures**: capture the error message, diagnose, retry with incremental changes, and report the root cause if unresolved.

---

## Operating Principles
- Prefer MCP + notebook tools for all actions (instrument discovery to measurement to analysis, and revise measurement if needed).
- Check **parameter limits/units** before setting or sweeping any parameter.
- Keep measurement code **reproducible** in the notebook.
## Workflow

### 1) Environment & Instrument Discovery
1. `mcp_list_resources(detailed=false)` -- enumerate available MCP resources/tools.
2. `qcodes_instrument_info(*)` -- list all connected instruments.
3. For each instrument you intend to use:
   - `qcodes_instrument_info(name)`
   - For each parameter you will set or scan: `qcodes_get_parameter_info(...)` to confirm **units, limits, get/set capability, and safety bounds**.

### 2) Measurement
4. Pull sweep templates/examples via `mcp_get_resource(...)` (sweep code examples/templates).
5. Use `notebook_*` functions to write MeasureIt code in notebook cells using `Sweep1D`, `Sweep2D`, etc. Sweep code templates can be obtained via `mcp_get_resource(...)`.
6. Run the cell with `notebook_execute_active_cell(...)`.
7. Wait/monitor completion with `measureit_wait_for_sweep(...)`.
   - If it errors: stop, diagnose, adjust, and retry. Never "assume" results.

### 3) Data Retrieval & Analysis
8. `database_list_experiments()` -- find the relevant experiment(s) and IDs.
9. `database_get_dataset_info(dataset_id, code_suggestion=True)` -- obtain loading code and run it in the notebook.
10. Analyze in notebook cells (plots + numerical results + uncertainties).
    - If inconclusive, refine the sweep design and repeat measurement + analysis.

---

## Final Response Format (Single JSON)
Return **one** JSON object in a code block containing:
- `summary`
- `measurements` (what sweeps were run + key settings)
- `datasets` (experiment_id + dataset_id(s))
- `results` (numerical results with units + uncertainties)
- `notes` (assumptions/limitations. Be verbose here.)
- `mcp_feedback_critical` (issues encountered; or "No difficulties encountered...")
- `mcp_feedback_soft` (suggestions to improve tool descriptions/results; or "No suggestions...")
\end{lstlisting}

\end{document}